\documentclass[journal]{IEEEtran} 

\usepackage{amsmath, amsfonts, amssymb, mathtools}
\usepackage{graphicx}
\usepackage{array}
\usepackage{textcomp}
\usepackage{stfloats}
\usepackage{url}
\usepackage{booktabs}
\usepackage{verbatim}
\usepackage{float}
\usepackage{color}
\usepackage{cite}

\usepackage{algorithm}
\usepackage{algpseudocode}

\usepackage{xcolor}
\definecolor{myblue}{RGB}{0,90,180}
\definecolor{mygreen}{RGB}{0,130,0}

\usepackage[font=footnotesize]{caption}
\bstctlcite{IEEEexample:BSTcontrol}
\usepackage{subcaption}
\usepackage{hyperref} 
\newtheorem{remark}{Remark}

\begin{document}
	
	\title{A Channel Knowledge Map-Driven Two-Stage  Coordinated User Scheduling in Multi-Cell 
    
    Massive MIMO Systems}
	
	\author{Jiayang Wan, \IEEEmembership{Graduate Student Member, IEEE}, 
	Hongwei Hou, \IEEEmembership{Graduate Student Member, IEEE}, 
    \\
	Jiawei Zhuang, \IEEEmembership{Graduate Student Member, IEEE},     
    Wenjin Wang, \IEEEmembership{
			Member, IEEE},
        Shi Jin, \IEEEmembership{Fellow, IEEE}
		
	 \thanks{
		Jiayang Wan, Hongwei Hou, Jiawei Zhuang and Wenjin Wang are with the National Mobile Communications Research
		Laboratory, Southeast University, Nanjing 210096, China, and also
		with Purple Mountain Laboratories, Nanjing 211100, China (e-mail:
		\{jywan, hongweihou, jw-zhuang, wangwj\}@seu.edu.cn).} 
	\thanks{
		 S. Jin is with the National Mobile Communications Research Laboratory, Southeast University, Nanjing 210096, China (e-mail:
		\{jinshi\}@seu.edu.cn).}	
	}

	\markboth{Journal of \LaTeX\ Class Files,~Vol.~14, No.~8, August~2021}%
	{Shell \MakeLowercase{\textit{et al.}}: A Sample Article Using IEEEtran.cls for IEEE Journals}

	\maketitle
	
	\begin{abstract}
		
In this paper, we investigate  narrowband coordinated user  scheduling in multi-cell  massive multiple-input multiple-output (MIMO) systems. 
Specifically, 
we formulate the  multi-cell coordinated user scheduling problem   under a spectral-efficiency maximization criterion, thereby revealing the inherent challenges in terms of 
\color{black}
computational complexity and signaling overhead.
To address these challenges, we  develop a user-scheduling-oriented CKM (US-CKM) for the first time and  further propose a US-CKM-driven two-stage coordinated user scheduling framework.
It exploits the mapping between location information and statistical channel state information (SCSI) to enable rapid retrieval and persistent reuse of SCSI, thereby substantially reducing the CSI acquisition overhead.
Moreover, by embedding statistical channel correlation into the CKM framework, it more effectively characterizes inter-user interference patterns.
Within this framework, we design an intra-cell active-user selection scheme for the first stage and an inter-cell coordinated scheduling scheme for the second stage, both based on the entries in the US-CKM.
Specifically, the first stage identifies users with favorable channel gains and low intra-cell interference, thereby substantially reducing the candidate user set while incurring only a marginal loss in sum rate.
The second stage, in turn, effectively suppresses inter-cell interference (ICI) by fully exploiting cross-cell channel correlations.                                           
To enhance robustness against imperfect SCSI in environments with dynamic scatterers,
\color{black}
 we augment the two-stage framework with a reliability-guided mechanism. 
Instead of treating all grids equally, we evaluate the stability of entries in the US-CKM using a grid reliability metric, which quantifies the variance of channel measurements at sampling locations within each grid. 
Based on this definition, we identify low-reliability grids, acquire their corresponding instantaneous CSI in real time, and integrate it with existing SCSI from other grids, thereby refining both channel gain and spatial correlation characteristics.
\color{black}
The proposed mechanism enables the system to strike an effective balance between   sum-rate performance and signaling overhead.
\color{black}
Simulation results confirm the effectiveness of the proposed framework, showing significant improvements in multi-cell sum-rate performance.
		
	\end{abstract}
	

\vspace{-5pt}

    \section{Introduction}\label{Introduction}

	\IEEEPARstart{W}{ith} the rapid development of wireless communication demands and increasing user density, multi-cell cooperative systems have emerged as a pivotal technology for enhancing network capacity and spectral efficiency \cite{xie2025multi,hou2024joint,hou2024tensor}. Through such  information exchange via backhaul links, the cooperation among base stations (BS) can significantly suppress  inter-cell interference (ICI), thereby improving cell-edge throughput and ensuring a more balanced quality of service \cite{hamza2013survey}.

	To fully harness the potential of multi-cell systems, 
	cooperative user scheduling plays a critical role, as the selection of scheduled users determines the interference characteristics, thereby constraining the achievable system throughput\cite{zhang2011weighted}. 
However, multi-cell scheduling faces two key challenges: its combinatorial nature leads to exponential complexity with the number of users \cite{sun2015beam,wang2024towards,wu2023low} and the reliance on instantaneous channel state information (ICSI) incurs significant signaling overhead \cite{hou2025tensor1}. These challenges highlight the urgent need for low-complexity scheduling algorithms for multi-cell systems.

\vspace{-8pt}

	\subsection{Prior Work}

	
	In the earliest years, the efforts of user scheduling predominantly targeted single-cell systems and evolved along two principal directions: spectral efficiency optimization \cite{dimic2005downlink,jiang2006greedy} and  computational complexity reduction \cite{yoo2006optimality,benmimoune2015joint,liu2015low,huang2012decremental,farsaei2019improved}.
	In studies of scheduling algorithms aimed at maximizing spectral efficiency, the predominant user selection algorithms are the greedy zero-forcing dirty-paper coding (DPC) algorithm \cite{dimic2005downlink} and the greedy scheduler with equal power allocation \cite{jiang2006greedy}. 
    However, both of these algorithms require sum-rate computation during scheduling, which introduces extremely high computational complexity and makes them difficult to implement in practical systems with a large number of antennas and users.
	To address this issue, the semi‐orthogonal user selection with zero‐forcing precoding (SUS‐ZF)\cite{yoo2006optimality} algorithm has been proposed. It selects a set of users with nearly orthogonal channels, facilitating ZF precoding while maintaining high spectral efficiency. Notably, when the number of users significantly exceeds the number of antennas, SUS‐ZF asymptotically achieves  the same sum‐rate performance as ZF‐DPC, but with substantially lower computational cost \cite{benmimoune2015joint}. 
    Apart from the SUS algorithm, several low-complexity user selection methods have been developed for multiple-input multiple-output (MIMO) systems. Liu et al. \cite{liu2015low} proposed a random selection algorithm, a straightforward approach without optimizing any metric. The delete-the-minimum-lambda (DML) method \cite{huang2012decremental} is a decremental selection based on ZF precoding, applicable only when the number of users is smaller than the number of antennas. The correlation-based selection (CBS) \cite{farsaei2019improved} maximizes the signal-to-interference-plus-noise ratio (SINR) of a particular user but does not guarantee globally optimal performance.

	When it comes to multi-cell systems, ICI becomes a vital bottleneck of system throughput, calling for multi-cell cooperation in user scheduling. To this end, several  cooperative user scheduling methods have been proposed in \cite{li2014multi,kwan2010survey,venturino2009coordinated,sawahashi2010coordinated}.
	A comprehensive overview of multi-cell coordinated scheduling and MIMO technologies in long term evolution (LTE) systems is provided in \cite{li2014multi}.
	Specifically, in the context of addressing ICI, uplink cooperative reception schemes are primarily divided into two categories: ICI suppression\cite{kwan2010survey}\cite{venturino2009coordinated} and ICI exploitation \cite{sawahashi2010coordinated,wang2024soft}. The ICI suppression approach assigns multiple users to the same resource block (RB) while adhering to predefined ICI thresholds, treating any residual interference as noise to simplify receiver design. In contrast, ICI exploitation adopts joint reception or detection techniques to transform interference into useful information, which leads to throughput improvement.
	Beyond interference mitigation itself,  the design of an effective cooperation architecture plays an equally vital role in achieving system-level performance gains. Specifically, centralized scheduling architectures  \cite{lahoud2016energy,sang2004coordinated,sun2013interference} treat all cooperating BSs as a virtual antenna array, jointly scheduling users across multiple  cells based on global channel state information (CSI). However, this  centralized strategy  imposes significant  demands on backhaul capacity \cite{lahoud2016energy} and centralized processing capabilities\cite{sang2004coordinated}, thus limiting its scalability in practical deployments   \cite{sun2013interference}. In contrast, distributed scheduling architectures \cite{cai2015joint} eliminate the need for a central controller, with each BS iteratively updating its local user set through limited information exchanges, thereby alleviating the burden on backhaul resources significantly.
    By incorporating advanced methods such as game-theoretic pricing and dynamic interference coordination algorithms \cite{cai2015joint}, distributed scheduling schemes can achieve performance comparable to centralized solutions, even under stringent signaling constraints.

	Despite their potential, most existing multi-cell cooperative scheduling schemes rely on the assumption  of perfectly up-to-date ICSI\cite{shan2024resource,cai2025cooperative}. This assumption, however, entails substantial signaling overhead due to frequent sounding reference signal (SRS) transmissions and significant performance degradation in  strong interference environments. To address this challenge, recent approaches have shifted toward reducing ICSI acquisition overhead by utilizing  long-term statistical CSI (SCSI)\cite{shi2020learning,ding2024improving,zhang2017sum}.
	In \cite{shi2020learning}, the authors leveraged the beam coupling characteristics of SCSI to formulate a beam-domain power control model and used a hybrid neural network to directly derive the ergodic rate from the channel covariance matrix (CCM) eigenvalue distribution, thus facilitating multi-cell scheduling. 
	The authors of \cite{ding2024improving} proposed a user grouping strategy for massive MIMO systems that groups users based on SCSI and improves system throughput.
	 Furthermore, \cite{zhang2017sum} introduced a joint beamforming and user scheduling scheme driven by SCSI, designed to optimize system performance and enhance scheduling efficiency, particularly in scenarios where acquiring ICSI is difficult or impractical. 
	More recently, the concept of  channel knowledge map (CKM)  has garnered extensive attention and numerous studies have investigated its construction and practical  applications \cite{zeng2021toward,wu2023environment,xu2024much}. By pre-mapping user geographic coordinates to long-term SCSI through offline processing, BSs can retrieve the required SCSI based solely on user location during real-time operation,  significantly reducing the CSI acquisition  overhead. Building upon this principle, the SCSI can serve as reliable prior information\cite{wu2023environment} to facilitate key operations such as beamforming, resource allocation and interference management\cite{xu2024much,wang2024robust,zhu2024joint}.

\subsection{Motivation and Main Contributions}

\color{black}

Recent advances in multi-cell scheduling have shown significant progress. However, three persistent challenges continue to hinder its practical scalability. The first challenge arises from the inherently combinatorial nature of user scheduling in a multi-cell environment, leading to computational complexity that grows exponentially with the number of users. The second is the reliance on accurate, real-time CSI, whose acquisition introduces substantial SRS overhead in large-scale systems. 
Third, centralized scheduling requires each BS to upload CSI to the central unit (CU), thereby incurring non-negligible information exchange overhead.

Motivated by the above analysis, this paper proposes a new type of CKM, termed the user-scheduling-oriented CKM (US-CKM), and accordingly develops a multi-cell coordinated user scheduling algorithm that maximizes spectral efficiency while reducing computational complexity, CSI acquisition overhead, and information exchange overhead.
The main contributions of this work are summarized as follows:

\begin{itemize}
\item With the objective of maximizing system throughput, we formulate the coordinated user scheduling problem in a multi-cell multi-user MIMO system. 
By exploiting the spatial consistency of channels, we propose and develop the US-CKM for the first time, which exploits the mapping between user location and statistical CSI to enable fast  retrieval and persistent reuse of statistical CSI, thereby substantially reducing the CSI acquisition overhead.
Furthermore, inspired by the interference-suppression requirement in user scheduling, we introduce statistical channel correlations into the CKM framework for the first time, which enables effective characterization of interference among users.

	\item To reduce computational complexity, we propose a two-stage coordinated user scheduling framework driven by US-CKM. 
In the first stage,
	two scenario-specific intra-cell active user selection algorithms
	are designed to significantly shrink the candidate user set while incurring only marginal sum-rate performance loss, thereby substantially reducing both information exchange overhead and computational complexity.
To mitigate the ICI, the second stage employs an
	inter-cell coordinated scheduling algorithm inspired by the
	SUS algorithm, which takes into account both channel
	gains and correlations and avoids the calculation of the
	receive beamformers.

	\item 	To enhance robustness against imperfect SCSI in environments with dynamic scatterers, we propose a robustness-enhanced version of the two-stage scheduling algorithm. Specifically, we introduce a grid reliability metric that quantifies the stability of channel knowledge in each US-CKM grid by measuring the variance across different sampling points within that grid. This metric supports targeted real-time acquisition of ICSI in low-reliability grids and fuses it with existing SCSI from reliable regions, thereby enabling a controllable balance between   sum-rate performance and signaling overhead.

		\color{black}
	\end{itemize}
	
	\color{black}
	
	\textit{Organization:} The remainder of this paper is organized as follows. Section~\ref{2} introduces the system model for uplink multi-cell multi-user scheduling. Section~\ref{3} presents the two-stage scheduling algorithm within an SCSI framework. Section~\ref{4} presents the robustness-enhanced two-stage scheduling algorithm framework. Section~\ref{5} provides comprehensive simulation results and performance evaluations. Finally, Section~\ref{6} summarizes the key contributions and concludes the paper.
	
	\textit{Notation:} Throughout this paper, bold uppercase letters  denote matrices, and bold lowercase letters  denote vectors. The operators $(\cdot)^T$, $(\cdot)^*$, and $(\cdot)^H$ represent the transpose, complex conjugate, and conjugate transpose, respectively.  The  $\mathrm{diag}(\mathbf{x})$ represents a diagonal matrix with elements from the vector $\mathbf{x}$. Furthermore, $\mathbb{E}\{\cdot\}$ denotes statistical expectation, and $\mathbb{C}^{m\times n}$ denotes the set of $m\times n$ complex-valued matrices. $\|\cdot\|_F$ and $\|\cdot\|_2$ denote the Frobenius norm and $\ell_2$ norm, respectively.

	\vspace{-5pt}

	\section{System Model and Problem Formulation}\label{2}
    
	In this section, we first introduce the received signal model for multi-cell multi-user systems. Following this model, the multi-cell scheduling problem is formulated based on the sum-rate maximization criterion. 
	
    \vspace{-5pt}
\subsection{Signal Model}

\begin{figure}[t]
		\centering		\includegraphics[width=3in]{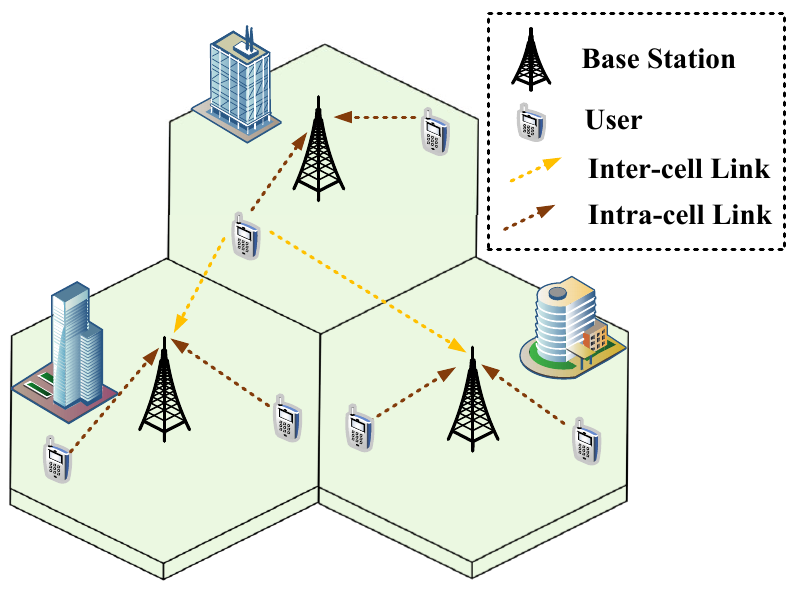}
		\caption{\small Illustration of the multi-cell uplink MIMO system.}
		\label{fig1}
	\end{figure}

	As illustrated in Fig. \ref{fig1}, we consider the uplink transmission of a multi-cell multi-user massive MIMO system consisting of \(L\) cells, where each cell is equipped with a  BS and $K$ randomly distributed single-antenna users \footnote[1]{Although the number of users may vary across cells, we assume, without loss of generality, that each cell contains the same number of users \cite{shi2020learning}  \cite{luo2024user}.}.
	The dual-polarized uniform planar array (DP-UPA) of each BS has $N = 2N_hN_v$ antennas, where $N_h$ and $N_v$ are the number of antennas in the horizontal and vertical dimensions, respectively.

Building on this system architecture, we further consider a cooperative scheduling framework. Specifically, we consider a system with $L$ coordinated cells that share a common pool of RBs, allowing users to dynamically form groups across cells. This coordination scheme is referred to as multi-cell cooperation, where a centralized controller is assumed to manage coordination among cells. 
Compared with single-cell approaches, multi-cell coordinated scheduling requires exchanging CSI and scheduling information between each cooperative BS and the central controller. In this paper, we assume cooperative BSs are connected to a centralized controller via high-speed backhaul links  with sufficient capacity for inter-cell coordination.

In this work, we consider  narrowband user scheduling.
 The centralized scheduler jointly schedules $L{\bar{K}}$
users among a total of $LK$ users to form a user group where $L{\bar{K}} \leq N$. Assuming that user group $\mathcal{U} = \{\mathcal{U}_{1}, \ldots, \mathcal{U}_{L}\}$ is scheduled, the received signal at the \(l\)-th BS, where $l \in \{1, \dots, L\}$, can be expressed as
	\vspace{-8pt}
    \begin{equation}
		\mathbf{y}_{l} = 
		\sum_{k \in \mathcal{U}_l}
		\mathbf{h}_{l,k} {x}_{l,k}+\sum_{\substack{m=1,m\neq l}}^{L} \sum_{k' \in \mathcal{U}_m} \mathbf{h}_{l,m,k'} {x}_{m,k'}  +  \mathbf{z}_l, 
		\label{eq:1}
	\end{equation}
	where $\mathcal{U}_l$ is defined as the set of scheduled users in the $l$-th cell, and  \({x}_{l,k}\) is the data symbol transmitted by the \(k\)-th user in the \(l\)-th cell, \(\mathbf{h}_{l,k} \in \mathbb{C}^{N \times 1}\)  and  \(\mathbf{h}_{l,m,k'} \in \mathbb{C}^{N \times 1}\) denote the channel  from the \(k\)-th user in the \(l\)-th cell and the \(k'\)-th user in \(m\)-th cell to the \(l\)-th BS, respectively, the additive white Gaussian noise at the $l$-th receiver is modeled as $\mathbf{z}_{l}\sim\mathcal{CN}\!\bigl(\mathbf{0},\sigma^{2}\mathbf{I}\bigr)$, and $\sigma^{2}$ denotes the noise power.
	Therefore, the uplink receiver  can be expressed as
    \vspace{-3pt}
	\begin{equation}
		{\hat{y}}_{l,k} =   \mathbf{w}_{l,k} ^{H} \mathbf{y}_{l},\label{eq:2}
	\end{equation}
    \!where  \(\mathbf{w}_{l,k}\in \mathbb{C}^{N \times 1}\) denotes the  receive beamformer   for the \(k\)-th user in the \(l\)-th cell, which is typically designed based on principles such as the matched filter (MF), zero-forcing (ZF), or minimum mean square error (MMSE)\cite{price2007communication}. ${\hat{y}}_{l,k}$  can be further decomposed into the sum of the desired signal, intra-cell interference, ICI, and noise, i.e.,
    \vspace{-3pt}
	\begin{align}
		{\hat{y}}_{l,k} = 
		& \underbrace{\mathbf{w}_{l,k}^\text{H}\mathbf{h}_{l,k} {x}_{l,k} }_{\text{Desired  signal}} \notag 
		+\! \underbrace{\sum_{\substack{i \neq k \\
					i \in \mathcal{U}_l  }} \!\mathbf{w}_{l,k}^\text{H}\mathbf{h}_{l,i} {x}_{l,i} }_{\text{Intra-cell interference}} \notag \\
		& + \!\underbrace{\sum_{\substack{m=1 \\ m\neq l}}^{L}   \sum_{k' \in \mathcal{U}_m} \mathbf{w}_{l,k}^\text{H}\mathbf{h}_{l,m,k'} {x}_{m,k'} }_{\text{ICI }}  +\underbrace{   \mathbf{w}_{l,k}^\text{H}\mathbf{z}_l}_{\text{Noise}}.	\label{eq:3}
	\end{align}
	
	With the assumption that \(\mathbb{E}[|{x}_{l,k}|^2] = 1\), the SINR  of the \(k\)-th user in the \(l\)-th cell is given by
    \vspace{-3pt}
	\begin{equation}
		\begin{split}
			\gamma_{l,k} ={}& 
			\frac{\displaystyle
				|\mathbf{w}_{l,k}^\text{H}\mathbf{h}_{l,k}|^2
			}{
				\displaystyle
				\sum\limits_{\substack{i \in \mathcal{U}_l\\i \neq k}}
				|\mathbf{w}_{l,k}^\text{H}\mathbf{h}_{l,i}|^2
				\!\!+\!\!\!
				\sum\limits_{\substack{m=1 \\m\neq l}}^{L}
				\sum\limits_{k' \in \mathcal{U}_m}\!\!\!
				|\mathbf{w}_{l,k}^\text{H}\mathbf{h}_{l,m,k'}|^2
				\!\!+\!\!
				\|\mathbf{w}_{l,k}^\text{H}\|^2\sigma^2
			}.
		\end{split}
		\label{eq:4}
	\end{equation}	
	
	It can be seen that intra-cell interference and ICI  will lead to significant degradation in SINR, which motivates us to design an effective scheduling algorithm to enhance the system throughput.
	
\vspace{-5pt}

	\subsection{Problem Formulation}
Assuming perfect knowledge of the ICSI, the multi-cell user scheduling  is formulated as the sum-rate maximization problem, expressed by
\vspace{-5pt}
\vspace{-5pt}
	\begin{align}
		\mathcal{P}_1:  \max_{\mathcal{U}_1...\mathcal{U}_L}\quad & R_{\text{sum}}= \sum_{l=1}^{L} \sum_{k\in \mathcal{U}_l}  
		\log_2  \!\left(1\! + \gamma_{l,k}	\right) , \label{eq:5}  \\ 
		{\rm   s.t.}  \quad  & |{\mathcal{U}_l}| = {\bar{K}}, \quad \forall l, \tag{5b}\label{eq:5b} \\
		& {\mathcal{U}_l \subset \mathcal{K}_l} \;, \quad \forall l,  \tag{5c}\label{eq:5c}
	\end{align}
    \color{black}
	where $\mathcal{K}_l$ is defined as the set of all users covered by the $l$-th BS. Let $\mathcal{K} = \{\mathcal{K}_1, \ldots, \mathcal{K}_L\}$ denote the collection of all user sets, and let ${\bar{K}}$ represent the number of users scheduled in each cell.
	To facilitate the design of multi-cell scheduling algorithms, we explicitly analyze two fundamental challenges that arise from problem $\mathcal{P}_1$:
	\begin{itemize}
		\item The optimization variables of $\mathcal{P}_{1}$ are the collection of sets $\mathcal{U}_l$ for $l = 1, \dots, L$, making this a typical combinatorial optimization problem, which is generally intractable in polynomial time.
       Although a greedy algorithm offers a direct solution, its prohibitive computational complexity limits its applicability in practice.

		\item The evaluation of SINRs in the objective function requires accurate ICSI for all users, which is typically obtained through   SRS. However, the fundamental mismatch between limited SRS resources and the large number of users in multiple cells presents significant challenges for CSI acquisition, thereby undermining the feasibility of user scheduling algorithms based on the above formulation.
	\end{itemize}
	
	These challenges motivate the development of two specific methodologies to address their corresponding issues:
	\begin{itemize}
		
		\item To address the first challenge, it can be observed that the scheduled user set with high system throughput typically consists of users with favorable channel gains and low mutual interference. This suggests that users with poor effective channel quality or high interference contributions can be excluded from the candidate set prior to scheduling.  By narrowing down the user pool in this manner, the computational complexity of the scheduling process can be substantially reduced with minimal impact on overall performance.

		\item 	In response to the second challenge, we observe that  the wireless channel remains approximately quasi-static in  propagation environments with static scatterers, such that the channel of each user is predominantly determined by its geographic location. This enables the construction of US-CKM, where the location serves as the query key for CSI retrieval, thereby eliminating the need for frequent SRS transmission. Moreover, as SCSI varies more slowly than the ICSI, the use of SCSI in US-CKM construction can further reduce the update frequency and provide a more stable foundation for user scheduling.

	\end{itemize}
	The specific algorithmic framework designed to address these challenges will be described in detail in the next section.

	\section{Two-Stage Scheduling Algorithm via SCSI}\label{3}

	In this section, we propose a grid-based US-CKM by exploiting the spatial consistency of channels  that incorporates both individual user characteristics and inter-user statistical properties. Building upon this, we present a two-stage user scheduling algorithm via SCSI, where the first stage targets intra-cell user selection, and the second stage handles ICI. 
	
\vspace{-5pt}

	\subsection{Grid-Based US-CKM Construction}\label{3-1}

	Before introducing the proposed grid-based US-CKM construction, we briefly revisit the characteristics of wireless communication channels. 
	In the context of the considered multi-cell multi-user massive MIMO  system, with the BS location fixed, the variation of channel $\mathbf{h}[t]$ mainly arises from two sources:  the variation in  user positions, denoted as $\mathbf{q}[t]$, as well as the variation of the actual communication propagation,  denoted as $E[t]$ \cite{wu2023environment}. Therefore, a generic representation of the  channel   can be expressed  as
    \vspace{-5pt}
	\begin{equation}
		\mathbf{h}[t] = f\bigl(\mathbf{q}[t], E[t]\bigr),
	\end{equation}
	where \(f(\cdot,\cdot)\) is an arbitrary function that maps from location and environment to channel knowledge. 
With this representation, accurately characterizing \(f(\cdot,\cdot)\) in complex environments remains extremely challenging, primarily due to the difficulty of mathematically modeling the environment \(E[t]\) and capturing the intricate interactions between radio waves and their surroundings.
	\color{black}
 Fortunately, this issue can be effectively alleviated by leveraging the CKM concept in \cite{wu2023environment}. Under the viewpoint widely adopted in the CKM literature \cite{wu2023environment,xu2024much,Zeng2024}, the slowly varying geometry and large-scale statistical characteristics dominated by static structures such as buildings and walls can be approximated as piecewise quasi-static within a CKM validity interval, whereas blockage or reflection variations and the resulting channel transitions induced by mobile objects such as vehicles, trucks, and pedestrians are inevitable.
	\color{black}

\begin{figure*}[t]
		\centering
		\includegraphics[width=6.5
        in]{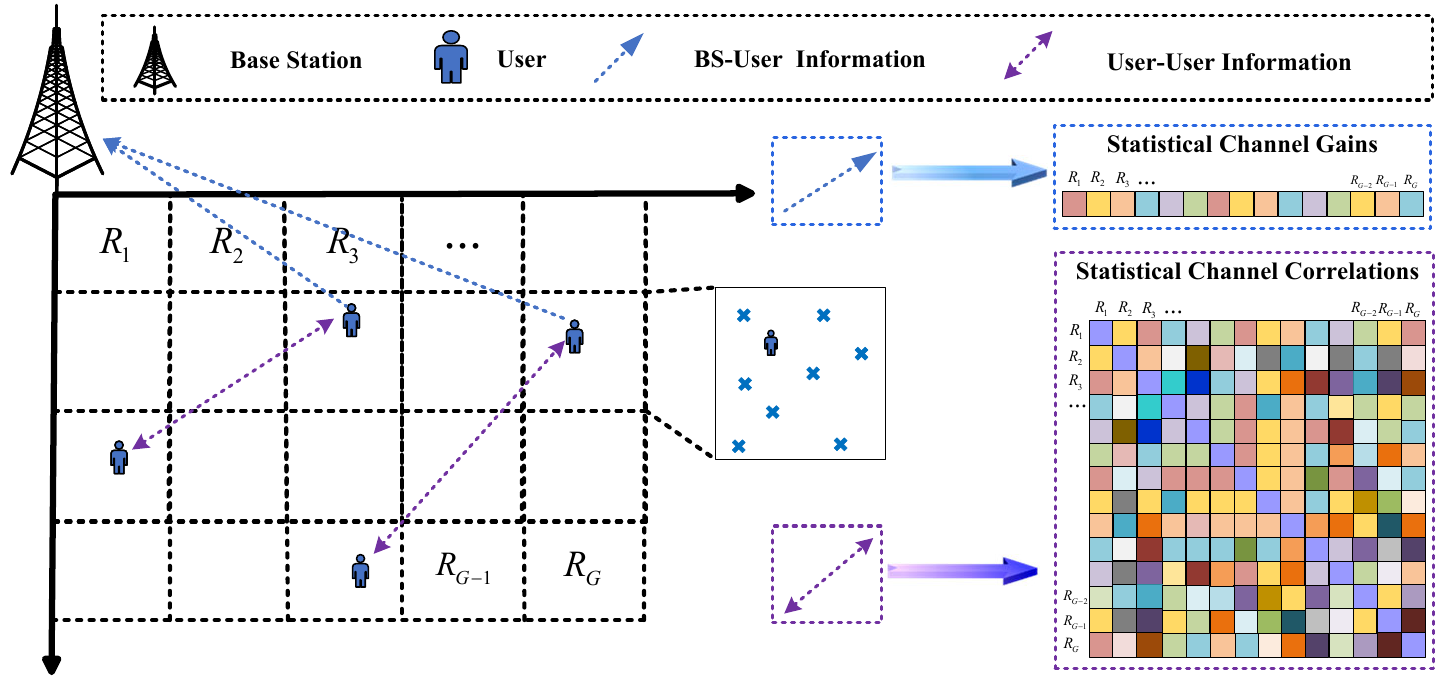}
		\caption{\small Illustration of grid-based US-CKM. The upper right is a one-dimensional table representing statistical channel gains obtained for all grids, and the lower right is a two-dimensional table representing statistical channel correlations between all grids.}
		\label{fig2}
	\end{figure*}

    Nevertheless, traditional CKMs suffer from severe storage challenges, as they need to store the CSI of every individual location.
	To address the storage issues, we further introduce a grid-based approximation by leveraging the spatial consistency of wireless channels, which divides the coverage area of each BS into grids and assumes that all locations within each grid share similar propagation environments. 
	Building on this, we shift from storing the channel mapping at every infinitesimal location to maintaining it only for a finite number of grids, with the grid size controlling the trade-off between storage accuracy and overhead, as shown in Fig. \ref{fig2}.

Under the grid-based approximation, it is necessary to identify the CSI types required for scheduling. In multi-cell multi-user systems, user scheduling relies on selecting candidates with favorable channels while minimizing inter-user interference—implying that CSI must capture both link quality and  correlations across users.

    \color{black}
Inspired by the channel gain map (CGM) \cite{xu2024much}, which is a site-specific database obtained via empirical field measurements or ray tracing simulations, we use the channel power gain to quantify the channel quality between each user and its serving BS, thereby facilitating the scheduling of users with favorable channel conditions.
    \color{black}
	For a specific grid, we obtain the statistical channel gain   of that grid  using various sampling methods \cite{xu2024much} and by averaging the channel gain of all sampling points. 
    \color{black}
    Specifically, the statistical channel gain  $\epsilon_{l,m,g}$ from the \(g\)-th grid in the \(m\)-th cell to the \(l\)-th BS  can be determined as 
    \vspace{-5pt}
\begin{equation}
	\epsilon_{l,m,g} = \frac{1}{S}\sum_{s=1}^{S}\left\|\mathbf{h}_{l,m,g,s}\right\|^2,
	\label{eq:7}
\end{equation}
        \color{black}
	where $\mathbf{h}_{l,m,g,s} \in \mathbb{C}^{N \times 1}$ represents the spatial domain channel from the $s$-th sampling point of the $g$-th grid in the $m$-th cell to the $l$-th BS.	

\color{black}
\begin{remark}\label{remark1}
	\textit{In this paper, we consider a centralized uplink multi-cell coordination architecture \cite{jafri2022robust}, where the cross-cell channels are acquired using an uplink SRS based listening and estimation mechanism. The central unit (CU) coordinates the SRS resources within the coordination cluster and exchanges with each BS only the minimal required control information over the backhaul/fronthaul, including the SRS time-frequency resource locations, sequence/port identifiers, and the necessary power-control parameters \cite{GiordanoWCNC2018SRScoord}. Once the BS has acquired this information, the cross-cell channel acquisition can be reformulated as a conventional pilot-based channel estimation problem, and the cross-cell channels can be readily obtained using well-established  algorithms \cite{hou2024joint,zhuang2025dmrs}.}
\end{remark}
\color{black}

	In user scheduling,  channel correlations are as critical as channel quality, since they determine the interference level and thus the achievable sum rate.
	Towards this end, we propose the concept of statistical channel correlations in the US-CKM  framework to describe the average channel correlation between the grids, which is based on the statistical channel.
	Specifically, under the reception of the \(l\)-th BS, the statistical channel correlation $\bar{\rho}_{l,m,g,q,b}$ between the \(g\)-th grid in the \(m\)-th cell and the \(b\)-th grid in the \(q\)-th cell can be expressed as
        \vspace{-5pt}
	\begin{equation}
		\bar{\rho}_{l,m,g,q,b} = \frac{\left|{\bar{\mathbf{h}}}_{l,m,g}^\text{H} {\bar{\mathbf{h}}}_{l,q,b}\right|}{\left\|{\bar{\mathbf{h}}}_{l,m,g}\right\| \times \left\|{\bar{\mathbf{h}}}_{l,q,b}\right\|},
		\label{eq:8}
	\end{equation}
	where $\bar{\mathbf{h}}_{l,m,g} = \frac{1}{S}\sum_{s=1}^{S}\mathbf{h}_{l,m,g,s}$ denotes the statistical channel from the \(g\)-th grid in the \(m\)-th cell to the \(l\)-th BS.
    \color{black}
For notational convenience, we define the statistical channel gain from the $g$-th grid in the $l$-th cell to the $l$-th BS as $\epsilon_{l,g}$, and denote by $\bar{\rho}_{l,g,g'}$ the statistical channel correlation between the $g$-th grid and the $g'$-th grid in the $l$-th cell.
\color{black}

Unlike traditional CKM approaches, in our proposed US-CKM, each US-CKM entry stores not only the long-term channel statistics between a specific grid and its serving BS, but also explicitly captures the critical statistical channel correlations across different grids.
Consequently, the US-CKM provides a reliable basis for subsequent scheduling algorithms, facilitating more accurate user scheduling decisions.

\begin{remark}\label{remark3}
\textit{Since a propagation environment typically consists of dominant static scatterers (e.g., buildings and walls) together with a small portion of dynamic scatterers (e.g., vehicles and pedestrians), the wireless channel can be decomposed into deterministic components contributed by the dominant static scatterers and small-scale random components induced by the small portion of dynamic scatterers \cite{jaeckel2018efficient}. By statistically averaging the channel realizations within a given grid, the resulting statistical channel $\bar{\mathbf{h}}$ suppresses small-scale random fluctuations related to the sampling point while retaining the long-term stable component determined by the dominant grid-related geometry \cite{zhu20213gpp}. Therefore, within a scheduling window or a CKM validity period, the corresponding statistical correlation coefficient $\bar{\rho}$ also exhibits temporal stability.}
\end{remark}

By constructing the US-CKM, the  real-time  acquisition of ICSI for a large number of users within each cell is transformed into obtaining the CSI of a limited set of grid-sampled users during idle periods, thereby alleviating the bottleneck imposed by limited SRS resources on user scheduling. Building on the US-CKM in each cell,  we present a two-stage user scheduling algorithm, where the first and second stages handle intra-cell user scheduling and ICI coordination. 

    \vspace{-5pt}

	\subsection{Stage~I: Intra-Cell Active User Selection}\label{3-2}

	In the first stage, we select a subset of active users with favorable channel quality and low intra-cell interference, thereby effectively reducing the problem scale. Toward this end, two active user selection algorithms are introduced: the Alternating Elimination and Selection (AES)   and the Global Interference Selection (GIS) algorithms. Specifically, AES employs a straightforward eliminate-and-select strategy, making it efficient in weak‐interference scenarios, while GIS computes cumulative inter‐user correlation each iteration to optimize scheduling under strong interference.

	\subsubsection{AES Algorithm}

	The exhaustive user-set enumeration in multi-cell multi-user massive MIMO system is computationally intractable, whereas the naive gain-based heuristics will overlook inter-user correlation and leave
	strong residual interference. 
	To address this limitation, the AES algorithm jointly considers channel gains and inter-user correlations to iteratively construct a user set featuring favorable channel quality and approximate orthogonality between users’ channels including the selection and elimination steps.
    \color{black}
	Specifically, the selection step identifies  the user with the maximum  channel gain from the current candidate pool 
       \color{black}
    and adds it to the active set, while  the elimination step removes  any remaining candidate whose channel correlation with any selected user exceeds a preset threshold to ensure  semi-orthogonality.
\color{black}
These two steps are iteratively performed until $K'$ users are selected from the $K$ candidate users in each cell, where $K$ denotes the number of users per cell and $K'$ denotes the number of users retained after the active user selection  phase per cell.  This process yields a user set that simultaneously exhibits favorable channel conditions and low mutual interference. The detailed active user selection procedure is summarized in Algorithm \ref{alg:AES}, where $\alpha$ denotes the correlation threshold parameter used in the AES screening process; the algorithm is executed sequentially in each cell.
\color{black}

\vspace{-1pt}
	
	\begin{algorithm}[t]
		\caption{Alternating Elimination and Selection}
		\label{alg:AES}
		\begin{algorithmic}[1]
			\State \textbf{Input:} $\mathcal{K}_l$, $K'$, ${\alpha}$,
            $\epsilon_{l,g}$,
            $\bar{\rho}_{l,g,g'}$,
             user coordinates $(x_k,y_k)$;

			\State Initialize: $\mathcal{A}_l = \varnothing$,
			$i = 1$, $\mathcal{T}_i=\mathcal{K}_l$;

            \State   For each $k\!\in\!\mathcal{K}_l$, find its grid index $g(k)$ based on $(x_k,y_k)$;
        
            \State For each $k\in\mathcal{K}_l$, the user-based ICSI is replaced by the grid-based SCSI:     
$\epsilon_{l,k} \leftarrow \epsilon_{l,g(k)}$ and 
$\bar{\rho}_{l,k,k'} \leftarrow \bar{\rho}_{l,g(k),g(k')}$;

            \While{$|\mathcal{A}_l| < K' $}
			\State Find the $i$-th user as follows:
			$\displaystyle \pi(i) = \arg\max_{k \in \mathcal{T}_i} \epsilon_{l,k}$;
			\State $\mathcal{A}_l = \mathcal{A}_l \cup \{\pi(i)\}$;
			\State   $\mathcal{T}_i = \mathcal{T}_i \setminus \{\pi(i)\}$;
			\If{$|\mathcal{A}_l| < K' $}
			\State 
			$\mathcal{T}_{i+1}
			= \bigl\{\,k \in \mathcal{T}_i \,\bigm|\, \forall k' \in \mathcal{A}_l,\;\bar{\rho}_{l,k,k'} \le {\alpha} \bigr\}$;
			\State $i = i + 1$;
			\EndIf
			\EndWhile
            \State \textbf{Output:}  $\mathcal{A}_l$.
		\end{algorithmic}
	\end{algorithm}
	It is noted that \({\alpha}\) and \({K'}\) play a crucial role in the trade-off between
	interference mitigation and throughput maximization: an excessively large ${\alpha}$ results in high residual interference, while a sufficiently small value may limit the number of users that can be selected.  Similarly, an undersized  $K'$ carries the risk of  excluding favorable-quality users, whereas an oversized one  may introduce excessive interference. 
	
		\begin{algorithm}[t]
		\caption{Global Interference Selection}
		\label{alg:GIS}
		\begin{algorithmic}[1]
			\State \textbf{Input:} $\mathcal{K}_l$, ${K'}$, 
			$\bar{\rho}_{l,g,g'}$,
			user coordinates $(x_k,y_k)$;
			
			\State Initialize: $\mathcal{A}_l = \varnothing$,
			$i = 1$, $\mathcal{T}_i=\mathcal{K}_l$;
			
			\State   For each $k\!\in\!\mathcal{K}_l$, find its grid index $g(k)$ based on $(x_k,y_k)$;
			
			\State For each $k\in\mathcal{K}_l$, the user-based ICSI is replaced by the grid-based SCSI: 
			$\bar{\rho}_{l,k,k'} \leftarrow \bar{\rho}_{l,g(k),g(k')}$;
			\While{$|\mathcal{T}_i| > {K'}$}
			\ForAll{$k \in \mathcal{T}_i$}    
			\State $\displaystyle \zeta_{l,k} = \sum_{k' \in \mathcal{T}_i} {\bar{\rho}}_{l,k,k'}$;
			\EndFor   
			\State Find the $i$-th user as follows:
			
			$\pi(i) = \arg\max_{k \in \mathcal{T}_i} \zeta_{l,k}$;
			\State Delete the $i$-th user: $\mathcal{T}_{i+1} = \mathcal{T}_i \setminus \{\pi(i)\}$;
			\State $i = i + 1$;
			\EndWhile
			\State \textbf{Output:}  $\mathcal{A}_l=\mathcal{T}_i.$
		\end{algorithmic}
	\end{algorithm}
	
	\color{black}

	Despite the advantages of algorithmic simplicity and low computational complexity, the AES algorithm faces the significant limitation of poor performance in high-interference environments. In each user selection, AES solely considers the interference imposed by a candidate user on the set of already selected users. It does not account for the interference said user might inflict upon the remaining pool of unselected users. Consequently, while AES performs localized interference optimization at each iteration, it fails to comprehensively incorporate the impact of global interference dynamics on subsequent user selection decisions.

	\subsubsection{GIS Algorithm}\label{3-3-2}
	To overcome the limitation of the AES algorithm, we propose the GIS algorithm that is explicitly designed to minimize overall interference throughout the  user selection process.
	The core idea of the  GIS algorithm is to preliminarily select a subset of candidate users based on their channel correlations to the final user selection. Unlike the AES algorithm, the GIS algorithm introduces the novel metric termed global interference metric,  which quantifies the cumulative interference a user causes to all other users within the same cell, thus providing a more holistic interference-aware selection criterion. The detailed procedure of the GIS algorithm is described in Algorithm \ref{alg:GIS}.

	The comparative analysis between the AES and GIS algorithms reveals their respective strengths and trade-offs. 
	Specifically, the AES algorithm makes each scheduling decision using only the interference among the already-selected users, 
	resulting in minimal complexity and making it suitable for lightly loaded, low-interference systems.
	In contrast, the  GIS algorithm 
	recalculates a global interference metric at each iteration, 
	capturing  the impact of each candidate on both scheduled and unscheduled users. This broader consideration lowers the mutual interference in the final selected user set, particularly benefiting  dense or large-scale deployments. However, the need to recompute the global interference metric at each iteration increases the computational complexity of GIS algorithm compared to the AES algorithm. This conclusion is further validated by the simulation results in Section \ref{5}.

	Based on the AES and GIS  active user selection algorithms, a set of active users \(\mathcal{A}_l\) has been effectively identified for each cell \(l\), where the selected user channels \(\bar{\mathbf{h}}_{l,1}, \bar{\mathbf{h}}_{l,2}, \ldots, \bar{\mathbf{h}}_{l,K'}\) maintain favorable channel gains while exhibiting low intra-cell interference.
	Therefore,   user scheduling can be executed directly on the active user set without significant performance degradation, thereby reducing the original optimization problem to the following simplified form
    \vspace{-5pt}
	\begin{align}
		\mathcal{P}_2: \max_{\mathcal{U}_1...\mathcal{U}_L}\quad & R_\text{sum}= \sum_{l=1}^{L} \sum_{k\in \mathcal{U}_l}  
		\log_2  \!\left(1\! + \gamma_{l,k}	\right) , \label{eq:17}  \\ 
		{\rm   s.t.}  \quad  & |{\mathcal{U}_l}| = {\bar{K}}, \quad \forall l, \tag{9b}\label{eq:9b} \\
		& {\mathcal{U}_l \subset \mathcal{A}_l} , \quad \forall l.  \tag{9c}\label{eq:9c}
	\end{align}
	
	\color{black}
	In this manner, the original problem is reduced to selecting users from the active-user set to maximize the system sum-rate.   
	By substantially shrinking the candidate user pool, the proposed algorithm significantly reduces both the problem dimension and the computational complexity, while incurring only a marginal loss in sum-rate performance relative to the greedy algorithm.   
	In the next subsection, we introduce a new scheduling algorithm to further suppress ICI more effectively, thereby approaching near-optimal user-selection performance. 
	\color{black}

\subsection{Stage~II: Inter-Cell Coordinated Scheduling}\label{3-3}
	
	In a multi-cell multi-user massive MIMO system, the greedy user scheduling algorithm that maximizes the system sum-rate requires recalculating the receive beamformers for all candidate users at each iteration, resulting in significant computational burden. In contrast, the SUS algorithm directly focuses on channel correlation and does not involve the computation of the receive beamformers, offering advantages in terms of  computational complexity and spectral efficiency in single-cell scenarios. However, in a multi-cell environment, users suffer from not only intra-cell interference but also ICI from neighboring cells. Since SUS constructs the orthogonal basis solely from intra-cell channels, it cannot be directly applied to the multi-cell scenario considered in this work. This limitation motivates the development of a low-complexity scheduling algorithm specifically tailored to such systems.

\begin{algorithm}[t]
		\caption{Inter-Cell Coordinated Scheduling }
		\label{alg:ICGCS}
		\begin{algorithmic}[1]
			\State \textbf{Input:} 
				$\mathcal{A}$, 
            ${\bar{K}}$,  
            $\epsilon_{l,m,g}$,
        $\bar{\rho}_{l,m,g,q,b}$,
             user coordinates $(x_k,y_k)$;
			\State Initialize: $i = 1$, $l = 1$, $\mathcal{U}_l = \varnothing$, $\mathcal{T}_l(i)=\mathcal{A}_l$;

          \State   For each $k\in\mathcal{A}$, find its grid index $g(k)$ based on $(x_k,y_k)$;
        
            \State For each $k\in\mathcal{A}$, the user-based ICSI is replaced by the grid-based SCSI \eqref{eq:118};
            
			\For{$i = 1$ \textbf{to} ${\bar{K}}$}   
			\For{$l = 1$ \textbf{to} $L$}                                  
			\ForAll{$k \in \mathcal{T}_l(i)$}                   
			\State Compute  $\mu_{l,k}$ from \eqref{eq:metric};
			\EndFor
			\State 
			$	\pi(l,i) \;=\; \arg\max_{k \in \mathcal{T}_l(i)} |\mathbf{\mu}_{l,k}|;$ 
			\State $\mathcal{U}_l = \mathcal{U}_l \cup \{\pi(l,i)\}$;
			\State $\mathcal{T}_l(i) = \mathcal{T}_l(i) \setminus \{\pi(l,i)\}$;
			\EndFor
			\State $i=i+1$;
			\EndFor
            			\State \textbf{Output:} $\mathcal{U}$.
		\end{algorithmic}
	\end{algorithm}
    

	To circumvent the computation of 
the receive beamformers inherent in greedy algorithms, we propose a scheduling criterion based on the SUS framework that simultaneously considers both channel gains and correlations.
Since the initially selected active users exhibit low intra-cell correlation, resulting in approximately orthogonal channels, the process of constructing an explicit orthogonal basis required in conventional SUS can be relaxed without substantially compromising performance.
	 
	\subsubsection{Review of SUS algorithm}\label{4-2-1}

\color{black}
The SUS algorithm \cite{yoo2006optimality,benmimoune2015joint}
is widely used because it selects users whose channels are nearly orthogonal. 
\color{black}
 \color{black}
	Let $\mathcal{T}_l(i)$ denote the candidate set of unscheduled users at the $i$-th iteration in the $l$-th cell, with initialization $\mathcal{T}_l(1)$ including all users in cell $l$. Let $\mathcal{S}_l(i)$ denote the set of already selected users before iteration $i$. For each $k \in \mathcal{T}_l(i)$, the orthogonal residual vector is obtained via the Gram--Schmidt procedure with respect to the previously selected basis vectors $\{\mathbf{g}_{l,j}\}_{j=1}^{i-1}$ as
	\vspace{-5pt}
	\setcounter{equation}{9}
	\begin{equation}
		\mathbf{g}_{l,k}
		= \mathbf{h}_{l,k}
		- \sum_{j=1}^{i-1} \frac{\mathbf{h}_{l,k}^\mathrm{H} \mathbf{g}_{l,j}}{\|\mathbf{g}_{l,j}\|^2}\,\mathbf{g}_{l,j}.
	\end{equation}
\color{black}	
	where \(\mathbf{g}_{l,k}\) is orthogonal to the subspace spanned by the previously chosen basis vectors \(\{\mathbf{g}_{l,1},\dots,\mathbf{g}_{l,i-1}\}\). 
	Subsequently, the SUS algorithm selects the user with the maximum orthogonal component energy in the  \(l\)-th cell, given by
     \vspace{-5pt}
	\begin{equation}
		\pi(l,i) \;=\; \arg\max_{k \in \mathcal{T}_l(i)} \|\mathbf{g}_{l,k}\|.
	\end{equation} 
	
	
	Let $\mathcal{S}_l$ denote the set of users already selected in the $l$-th cell.  After such selection, we remove from the candidate set $\mathcal{T}_l(i)$ any user $k$ whose channel correlation with the newly added orthonormal basis vectors 
	$\{\mathbf{g}_{l,j} \mid 1 \le j \le |\mathcal{S}_l|\} $ exceeds a predefined threshold ${\alpha}$.
\color{black}	
	After selecting the $i$-th user $\pi(l,i)$, the candidate set is updated as
	\vspace{-5pt}
	\setcounter{equation}{11}
	\begin{equation}
		\mathcal{T}_l(i+1)
		= \left\{\,k \in \mathcal{T}_l(i)\setminus\{\pi(l,i)\}\;\middle|\;
		\frac{\left|\mathbf{h}_{l,k}^\mathrm{H} \,\mathbf{g}_{{l,i}}\right|}
		{\|\mathbf{h}_{l,k}\|\;\|\mathbf{g}_{{l,i}}\|} < \alpha
		\right\}.
	\end{equation}
	This update is repeatedly applied for each user $i$ in the set $S_l$, thereby ensuring that all users in the final selected set $\mathcal{S}_l$ are mutually semi-orthogonal.
\color{black}
However, SUS exhibits a key limitation in a multi-cell environment: since its semi-orthogonal basis vectors are constructed solely from single-cell channels, the selected users may still  suffer from significant ICI, thereby reducing system performance. 
	
	\subsubsection{SUS-inspired multi-cell user selection algorithm}\label{4-2-2}
	
	In the first stage, the AES/GIS algorithm  yields an active user set with sufficiently low intra-cell correlation, allowing each retained channel  to serve directly as a basis  and thereby eliminating the construction of orthogonal basis at every scheduling iteration.  
\color{black}
For each \(k \in \mathcal{T}_l(i)\), we extract the component of the channel vector \(\mathbf{h}_{l,m,k}\) that lies in the orthogonal complement of the subspace spanned by the previously selected channel vectors.
Specifically, define \(\mathcal{V}(i-1)\triangleq \mathrm{span}\big(\{\mathbf{h}_{l,q,j}\}_{q=1,\dots,L}^{j=1,\dots,i-1}\big)\),
where \(\mathbf{h}_{l,q,j}\) denotes the previously selected channel vectors used to construct the reference subspace \(\mathcal{V}(i-1)\).
Following the active-user selection in the first stage (Stage~I), these selected vectors are designed to be semi-orthogonal; hence, the projection onto \(\mathcal{V}(i-1)\) can be well approximated by the sum of individual projections onto \(\{\mathbf{h}_{l,q,j}\}\).
Then, \(\mathbf{h}^{\perp}_{l,m,k}\in \mathcal{V}(i-1)^\perp\) denotes the orthogonal residual of \(\mathbf{h}_{l,m,k}\) with respect to \(\mathcal{V}(i-1)\), obtained by subtracting its projection onto \(\mathcal{V}(i-1)\) as
\vspace{-5pt}
\begin{equation}
	{\mathbf{h}}_{l,m,k}^\perp 
	= {\mathbf{h}}_{l,m,k}
	- \sum_{q=1}^{L}\sum_{j=1}^{i-1}
	\frac{{\mathbf{h}}_{l,m,k}^\text{H}\,{\mathbf{h}}_{l,q,j}}
	{\|{\mathbf{h}}_{l,q,j}\|^2}\,
	{\mathbf{h}}_{l,q,j}.
	\label{eq:13}
\end{equation}
\color{black}
	where  it reduces to
${\mathbf{h}}_{l,m,k}^\perp = {\mathbf{h}}_{l,m,k}$ for $i = 1$. Combining \eqref{eq:8} and \eqref{eq:13}, the projection of the channel  can be expressed as
   \color{black}
\begin{equation}
	{\mathbf{h}}_{l,m,k}^{\perp}
	= {\mathbf{h}}_{l,m,k}
	- \sum_{q=1}^{L}\sum_{j=1}^{i-1}
	{\rho}_{l,m,k,q,j}\,
	\bigl\|{\mathbf{h}}_{l,m,k}\bigr\|\,
	{\hat{\mathbf{h}}}_{l,q,j},
	\label{eq:14}
\end{equation}
where $
\hat{\mathbf{h}}_{l,q,j} \!=\!
{\mathbf{h}}_{l,q,j}/\|{\mathbf{h}}_{l,q,j}\|$
is the unit direction of  \({\mathbf{h}}_{l,q,j}\)
and 
$\rho_{l,m,k,q,j}$ denotes the  channel correlation   between the $k$-th user   in the $m$-th cell and  the $j$-th user in the $q$-th cell under the reception of the \(l\)-th BS, which can be expressed as
$
\rho_{l,m,k,q,j}
=
\frac{\left|\mathbf{h}_{l,m,k}^{\mathrm H}\mathbf{h}_{l,q,j}\right|}
{\left\|\mathbf{h}_{l,m,k}\right\|\;\left\|\mathbf{h}_{l,q,j}\right\|}.
$
\color{black}
	Therefore, the  residual norm is expressed as
    \vspace{-5pt}
	\begin{equation}
		\|{\mathbf{h}}_{l,m,k}^{\perp}\|
		=\|{\mathbf{h}}_{l,m,k}\|\,
		\Bigl\|
		\hat{\mathbf{h}}_{l,m,k}
		-\!\!\sum_{q=1}^{L}\sum_{j=1}^{i-1}
		{\rho}_{l,m,k,q,j}\,\hat{\mathbf{h}}_{l,q,j}
		\Bigr\|.
		\label{eq:15}
	\end{equation}
	
	Since the previously selected $\{\hat{\mathbf h}_{l,q,j}\}_{q=1,\dots,L}^{j=1,\dots,i-1}$ are approximately unit-orthogonal vectors,  we have
    \vspace{-5pt}
	\begin{equation}
		\Bigl\|
		\hat{\mathbf{h}}_{l,m,k}
		-\!\!\sum_{q=1}^{L}\sum_{j=1}^{i-1}
		\rho_{l,m,k,q,j}\,\hat{\mathbf{h}}_{l,q,j}
		\Bigr\|^{2}
		= 1-\!\!\sum_{q=1}^{L}\sum_{j=1}^{i-1}
		\bigl|{\rho}_{l,m,k,q,j}\bigr|^{2},
		\label{eq:16}	
	\end{equation}
	where $\{\hat{\mathbf h}_{l,q,j}\}_{q=1,\dots,L}^{j=1,\dots,i-1}$ denotes the set of previously admitted unit-norm channels at the $l$-th BS, satisfying the approximate orthonormality condition. 
    A detailed step-by-step derivation of the resulting residual-energy expression is provided in Appendix \ref{APP}.
    By expanding the squared norm $\|\hat{\mathbf h}_{l,m,k} - \sum_{q=1}^{L}\sum_{j=1}^{i-1}{\rho}_{l,m,k,q,j}\,\hat{\mathbf h}_{l,q,j}\|^2$ and invoking the approximate orthogonality  to ignore cross-terms, one obtains the simplified expression in \eqref{eq:16}. Therefore, the multi-cell coordinated scheduling performance metric is
        \vspace{-5pt}
\begin{equation}
\mu_{l,k}
=
\sqrt{
\left\|\mathbf{h}_{l,m,k}\right\|^{2}
\left(
1-\sum_{q=1}^{L}\sum_{j=1}^{i-1}
\left|\rho_{l,m,k,q,j}\right|^{2}
\right)
}.
\label{eq:metric}
\end{equation}

	Leveraging the grid‐based US-CKM construction, we can employ grid‐based SCSI to develop a two-stage scheduling algorithm.
	\color{black}
	Specifically, for any candidate user $k$ located in the $g(k)$-th grid in the $m$-th cell and  user $j$ located in the $g(j)$-th grid  in the $q$-th cell, we first determine their corresponding grid indices $g(k)$ and $g(j)$ according to the user coordinates $(x_k,y_k)$ and $(x_j,y_j)$. Then, the channel gain  $\epsilon_{l,m,g(k)}$ and channel correlation  $\rho_{l,m,k,q,j}$ is approximated by 
	\setcounter{equation}{17}
	\begin{equation}
		\left\|\mathbf{h}_{l,m,k}\right\|^2 \leftarrow \epsilon_{l,m,g(k)}, \quad
		\rho_{l,m,k,q,j} \leftarrow \bar{\rho}_{l,m,g(k),q,g(j)},
		\label{eq:118}
	\end{equation}
	where $\epsilon_{l,m,g(k)}$ denotes  the statistical channel gain   from grid $g(k)$ in the $m$-th cell to the $l$-th BS and $\bar{\rho}_{l,m,g(k),q,g(j)}$ denotes the statistical channel correlation between grid $g(k)$ in the $m$-th cell and  grid $g(j)$ in the $q$-th cell under  the reception of the \(l\)-th BS.
	Consequently, the dependence on  ICSI is mitigated, thereby alleviating the SRS overhead.
	\color{black}

	In summary, the proposed criterion  extends SUS to the multi-cell system by embedding inter-cell correlations into the residual metric, thereby enabling coordinated user selection across cells. When inter-cell  correlation  is omitted, the criterion degenerates to the original SUS selection  without explicit orthogonal-basis construction. 
	The detailed inter-cell coordinated scheduling (ICCS) algorithm is summarized in Algorithm \ref{alg:ICGCS}. This  algorithm is executed sequentially with the user selection in each iteration influencing the scheduling decisions in subsequent iterations. By incorporating the inter-cell correlation,  the selected user set not only minimizes intra-cell interference but also reduces ICI, leading to  improved overall throughput performance.
	
    \color{black}
    

	\section{Robustness-Enhanced Two-Stage Scheduling Algorithm}\label{4}


Due to the errors introduced during various stages of CKM construction, the resulting portion of SCSI is rendered unreliable, which in turn degrades the scheduling performance.
To address this issue, we introduce a variance-based metric to evaluate SCSI reliability, and classify each grid as either reliable or unreliable. 
	By integrating both SCSI in the reliable grids and ICSI in the unreliable grids, we construct refined channel gain and correlation metrics to enhance the robustness of the proposed two-stage user scheduling algorithm.
	
	       \vspace{-5pt}
	\subsection{Statistical Prior Reliability Evaluation}\label{4-1}
	
\color{black}
SCSI unreliability is primarily driven by channel variations 
caused by the movement of dynamic scatterers, together with the 
accumulation of measurement noise, quantization errors, and 
interpolation errors throughout the map construction process
\cite{zeng2021toward}.
Therefore, it is crucial to distinguish grid points with unreliable SCSI from those with reliable SCSI, 
so that the unreliable grids can be handled separately.
\color{black}

When a grid exhibits significant variations between the geometric center and other sampling points, the grid-based SCSI may not accurately capture the actual SCSI of users.
In particular, the reliability of the grid-based SCSI is closely related to the fluctuation of the channel correlation within the grid, where larger fluctuations indicate lower reliability.
	To evaluate the reliability of each grid, the geometric center of each grid is  selected as a reference point, and then  the  channel correlation between any other position within the same grid and the reference point is measured.
	Specifically, under the reception of the \(l\)-th BS, the statistical correlation \(\rho^s_{l,m,g}\) between the $s$-th sampling point and the center in the \(g\)-th grid in the \(m\)-th cell can be expressed as
    \vspace{-1pt}
	\begin{equation}
		\rho^s_{l,m,g} = \frac{\left| {\mathbf{h}}_{l,m,g,s}^\text{H}  {\mathbf{h}}_{l,m,g,c} \right|}{\left\| {\mathbf{h}}_{l,m,g,s} \right\| \times \left\| {\mathbf{h}}_{l,m,g,c} \right\|}\;, \quad \forall s = 1, 2, \ldots, S
		\label{eq:18}
	\end{equation}
	where $\mathbf{h}_{l,m,g,s}$ and $\mathbf{h}_{l,m,g,c}$ denote the instantaneous channels  from the $s$-th sampling point and the center of the $g$-th grid in the $m$-th cell to the $l$-th BS, respectively.
	Building upon this, we define a variance-based metric $\sigma_{l,m,g}$  as the sample variance of the correlation coefficients. Specifically, $\sigma_{l,m,g}$, which is defined as the variance-based metric from the $g$-th grid in the $m$-th cell to the $l$-th BS, is expressed by		
        \vspace{-5pt}
\color{black}
\begin{equation}
	\sigma_{l,m,g}
	= \frac{1}{S}\sum_{s=1}^{S}\left(\rho_{l,m,g}^{s}-\frac{1}{S}\sum_{z=1}^{S}\rho_{l,m,g}^{z}\right)^{2},
	\label{eq:19_revised}
\end{equation}
where  \(\rho_{l,m,g}^{z}\)  represents the statistical correlation between  the $z$-th sampling point and the center in the \(g\)-th grid in the \(m\)-th cell under the reception of the \(l\)-th BS and  $S$ is the number of sampling points in each grid. 
\color{black}
	The reliability of the grid-based SCSI is determined by $\sigma_{l,m,g}$, where a larger $\sigma_{l,m,g}$ indicates greater channel fluctuation and lower reliability.

 \begin{algorithm}[t]
		\caption{Robustness-Enhanced Two-Stage Scheduling }
		\label{alg:4}
		\begin{algorithmic}[1]
			\State \textbf{Input:} $\mathcal{K}$, 
			${K'}$,  ${\bar{K}}$,
            grid-based SCSI from reliable grids, 
       user-based ICSI from unreliable grids;
			\State Initialize:  $\mathcal{U} = \varnothing$;
			\ForAll{$k $ $\in$ $\mathcal{K}$}  
			\State Compute  $\tilde{\mathbf{h}}_{l,m,k}$ from \eqref{eq:21};
			\EndFor
			\ForAll{$k $ $\in$ $\mathcal{K}$}  
			\State Compute  $\tilde{{h}}_{l,m,k}$ from \eqref{eq:22};
			\State Compute  $\tilde{\rho}_{l,m,k,q,k'}$ from \eqref{eq:23};
			\EndFor 
			\ForAll{$l = 1$ \textbf{to} $L$}  
			\State  Compute $\mathcal{A}_l$ via Algorithm~\ref{alg:AES} or Algorithm~\ref{alg:GIS}, 
       with the inputs replaced by $\tilde{h}_{l,k}$ and $\tilde{\rho}_{l,k,k'}$;
			\EndFor      	 
			\State Compute  $\mathcal{U}$ via  Algorithm \ref{alg:ICGCS}, 
       with the inputs replaced by  $\tilde{h}_{l,m,k}$ and $\tilde{\rho}_{l,m,k,q,k'}$; 
            \State \textbf{Output:} $\mathcal{U}$.
		\end{algorithmic}
	\end{algorithm}

	To facilitate  the practical application of the proposed  variance-based metric in scheduling, we introduce a predefined reliability threshold, which can classify each grid as either reliable or unreliable. This classification effectively partitions the US-CKM into two subsets: one containing grid-based SCSI that can be directly used for SCSI-based scheduling, and another requiring user-based ICSI acquisition. Specifically, we define the CSI reliability indicator for the 
	\(g\)-th grid in the 
	\(m\)-th cell with respect to the 
	\(l\)-th BS, given by
        \vspace{-5pt}
	\begin{equation}
		R_{l,m,g} =
		\begin{cases}
			1, & \sigma_{l,m,g} \le \delta, \\
			0, & \sigma_{l,m,g} > \delta, 
		\end{cases}
	\end{equation}
	where \(\delta\) is the reliability threshold. By tuning this parameter, the system can flexibly balance signaling overhead and statistical accuracy: lowering the threshold tightens the reliability criterion and improves statistical fidelity, while raising it relaxes the constraint and reduces the need for SRS-based measurements.

           \vspace{-5pt}

	\subsection{Robustness-Enhanced Two-Stage  Scheduling Algorithm}\label{4-2}

	To support more refined and reliable scheduling, both SCSI and ICSI are jointly exploited to compute the effective channel gain and inter-user correlation, depending on the reliability classification. 
	Under unreliable grid conditions, we update the SCSI with ICSI to obtain the more refined CSI. Specifically, the effective channel $\tilde{\mathbf{h}}_{l,m,k}$ from the $k$-th user of the $m$-th cell, located in the $g$-th grid, to the $l$-th BS is expressed as
           \vspace{-5pt}
	\begin{equation}
		\tilde{\mathbf{h}}_{l,m,k} =
		\begin{cases}
			\mathbf{h}_{l,m,k}, & \text{if } R_{l,m,g}=0, \\
			\bar{\mathbf{h}}_{l,m,g}, & \text{if } R_{l,m,g}=1,
		\end{cases}
		\label{eq:21}
	\end{equation}
where \(\mathbf{h}_{l,m,k}\) and \(\bar{\mathbf{h}}_{l,m,g}\) are the user‐based real-time channel  from the \(k\)-th user in the \(m\)-th cell to the \(l\)-th BS obtained via SRS‐based estimation and   the grid-based  statistical channel retrieved from the existing US-CKM, respectively.
	To achieve the coordinated user scheduling, we integrate both SCSI in the reliable grids and ICSI in the unreliable grids to construct refined channel gain and
	correlation metrics to enhance the robustness of the proposed two-stage user scheduling algorithm.
	Specifically, 
	the effective channel gain is defined as
           \vspace{-5pt}
	\begin{equation}
		\tilde{{h}}_{l,m,k} =
		\begin{cases}
			\left\|\mathbf{h}_{l,m,k}\right\|^2, & \text{if } R_{l,m,g}=0, \\
			\epsilon_{l,m,g}, & \text{if } R_{l,m,g}=1,
		\end{cases}
		\label{eq:22}
	\end{equation}
	where  \(\epsilon_{l,m,g}\) is the statistical channel gain from the \(g\)-th grid in the \(m\)-th cell to the \(l\)-th BS, as defined in \eqref{eq:7}.
	Furthermore, the effective channel correlation between	 the \(k\)-th user  in the \(g\)-th grid of the \(m\)-th cell and the \(k'\)-th user  in the \(b\)-th grid of the \(q\)-th cell, as observed from the \(l\)-th BS, is given by
           \vspace{-5pt}
	\begin{equation}
		\tilde{\rho}_{l,m,k,q,k'} =
		\displaystyle \frac{\left|\tilde{\mathbf{h}}_{l,m,k}^\text{H}\,\tilde{\mathbf{h}}_{l,q,k'}\right|}{\|\tilde{\mathbf{h}}_{l,m,k}\|\,\|\tilde{\mathbf{h}}_{l,q,k'}\|}.
		\label{eq:23}
	\end{equation}

	      \begin{figure}[t]
		\centering		\includegraphics[width=0.8\linewidth]{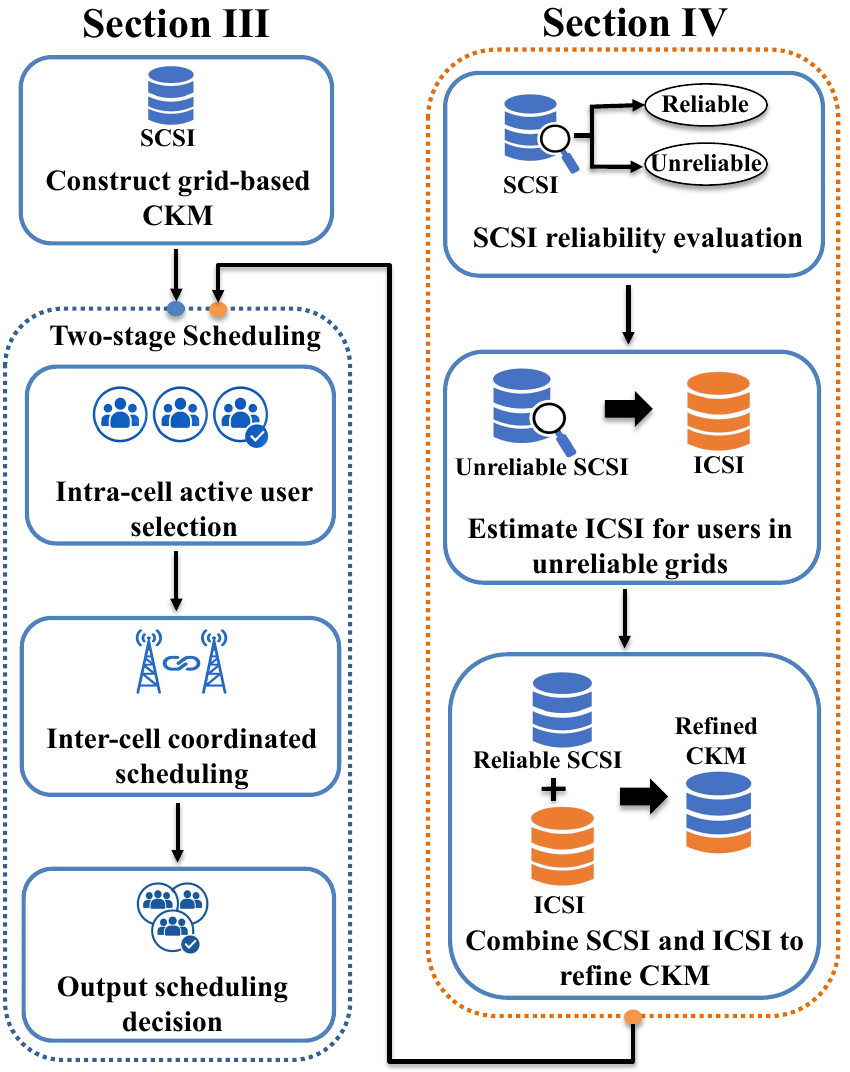}
		\caption{\small Flowchart of robustness-enhanced two-stage user scheduling algorithm.}
		\label{fig11}
	\end{figure}

By integrating both SCSI from the reliable grids and ICSI from the unreliable grids, the robustness of multi-cell user scheduling is significantly enhanced in propagation environments with static scatterers. Notably, this implicitly employs a two-level grid partitioning strategy, assigning finer grids to regions affected by static scatterers, while retaining the original grid size in statistically stable regions. This approach can be naturally generalized to a non-uniform grid design, where the grid resolution adapts to the local propagation characteristics, enabling a compact yet effective CSI representation and supporting efficient user scheduling.
\color{black}
With more effective CSI available, the inputs to Algorithms~\ref{alg:AES}, \ref{alg:GIS}, and \ref{alg:ICGCS} are no longer restricted to grid-based SCSI. Instead, they comprise the grid-based SCSI from reliable grids together with limited user-based ICSI from unreliable grids.
\color{black}

	In summary, the proposed  robustness-enhanced two-stage scheduling framework  consists of the following steps:
	\begin{itemize}
		\item \textbf{Reliability Determination:} Using the CSI reliability indicator, each grid’s CSI is classified as reliable or unreliable. 
        For unreliable grids, real-time scheduling of SRS is required to obtain ICSI.
		
		\item \textbf{Parameter Computation:} The effective channel gain and inter-user correlations are calculated from the obtained CSI, providing quantifiable inputs for subsequent scheduling.
		
		\item \textbf{Intra‐Cell User  Selection:} By applying Algorithm \ref{alg:AES} or \ref{alg:GIS}, intra‐cell interference is suppressed and an active user set is selected, substantially  narrowing the candidate set.
		
		\item \textbf{Inter‐Cell Coordinated Scheduling:} Algorithm \ref{alg:ICGCS} is subsequently  applied to the active user set, taking ICI into account to finalize user scheduling.
	\end{itemize}

	The corresponding algorithmic procedures are detailed in Algorithm \ref{alg:4} and the proposed  robustness-enhanced user scheduling framework is shown  in 
    Fig. \ref{fig11}.
	\color{black}
	To further clarify the merits of the proposed framework, its key features can be summarized from the following four aspects.
	\begin{itemize}
		\item \textbf{CSI Acquisition Overhead Reduction:} 
		It balances CSI accuracy and CSI acquisition overhead by selectively using historical SCSI and triggering SRS-based acquisition  in Section~\ref{4-2} only when necessary.

        \item \textbf{Information Exchange Overhead Reduction:} 
		By adopting a two-stage coordinated user scheduling framework in Section~\ref{3}, the algorithm first reduces the candidate user set and subsequently conducts inter-cell coordinated scheduling, thereby alleviating information exchange overhead and improving scalability for large-scale network deployments.

		\item \textbf{Robustness Under Imperfect Conditions:} 
		The proposed framework mainly builds on SCSI constructed offline and reused online, with instantaneous CSI updated on demand only for a small number of regions with unreliable SCSI in Section~\ref{4}. Consequently, feedback overhead is low, while the effects of fronthaul delay and packet loss remain limited \cite{jafri2024asynchronous}. Meanwhile, the slow variation of statistical CSI makes it less vulnerable to information staleness; for the few users requiring instantaneous CSI updates, channel prediction may be used to alleviate channel aging  \cite{wang2022channel}. Furthermore, ideal synchronization is assumed under centralized deployment \cite{bjornson2019making}.

		\item \textbf{Framework Extensibility:}
		The proposed framework can be further combined with robust beamforming methods \cite{jafri2024distributed} following user scheduling, thereby improving its robustness against CSI uncertainty. Moreover, although this study mainly concentrates on sum-rate maximization in \eqref{eq:5}, the framework can be naturally generalized to fairness-oriented designs through the incorporation of user-specific weights or additional scheduling constraints.
		
	\end{itemize}

\begin{figure*}[t] 
    \centering
    \begin{subfigure}[t]{0.31\textwidth}
        \centering
        \includegraphics[width=\linewidth]{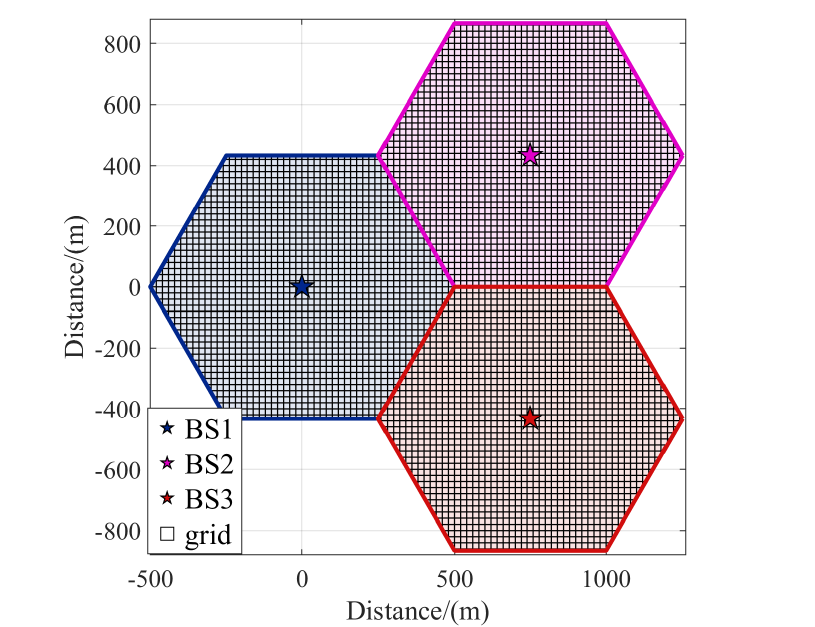}
        \caption{\small Grid partition under a three-cell layout.}
        \label{fig7.1}
    \end{subfigure}
    \hfill
    \begin{subfigure}[t]{0.31\textwidth}
        \centering
        \includegraphics[width=\linewidth]{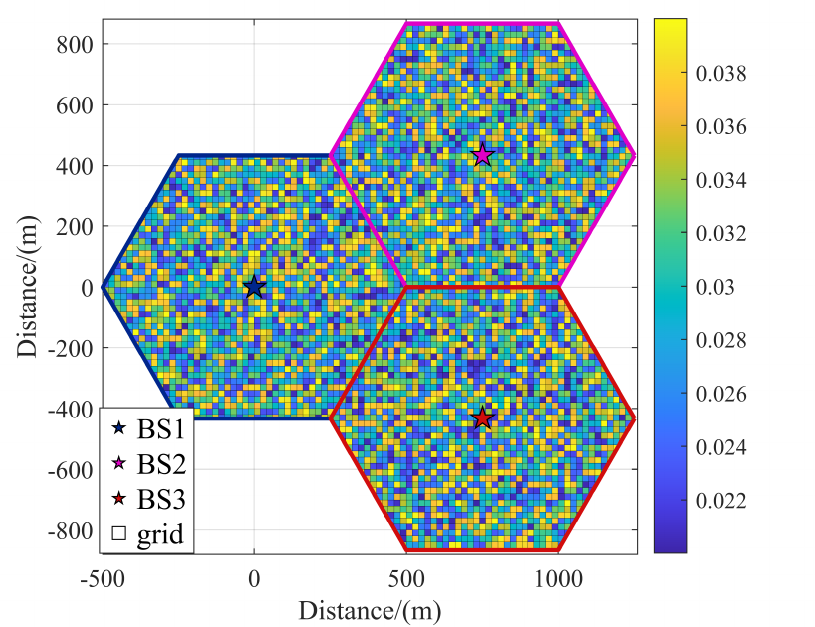}
        \caption{\small Statistical channel gain map.}
        \label{fig7.2}
    \end{subfigure}
    \hfill
    \begin{subfigure}[t]{0.31\textwidth}
        \centering
        \includegraphics[width=\linewidth]{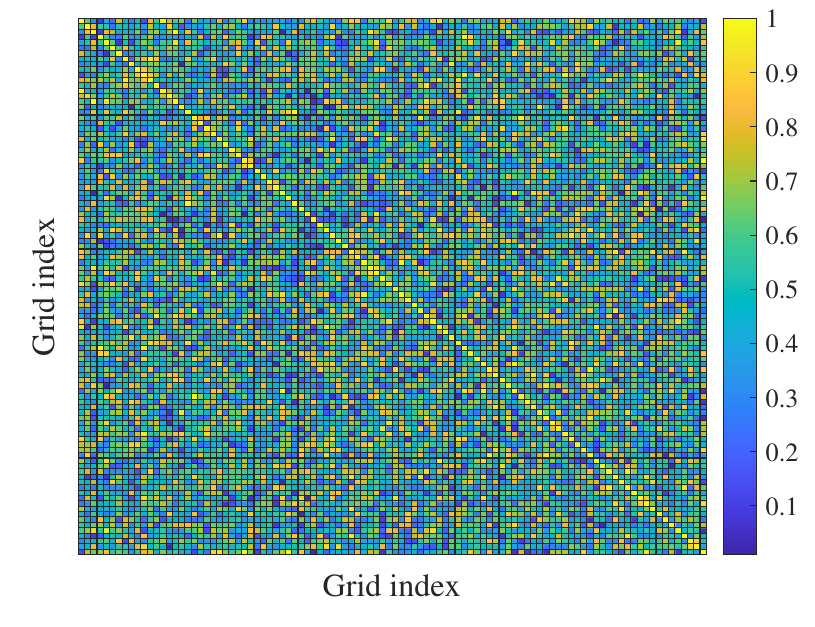}
        \caption{\small Statistical channel correlation map.}
        \label{fig7.3}
    \end{subfigure}

    \caption{Visualization of grid-based US-CKM.}
    \label{fig3333}
\end{figure*}

\begin{figure*}[t] 
    \centering
    \begin{subfigure}[t]{0.31\textwidth}
        \centering
        \includegraphics[width=\linewidth]{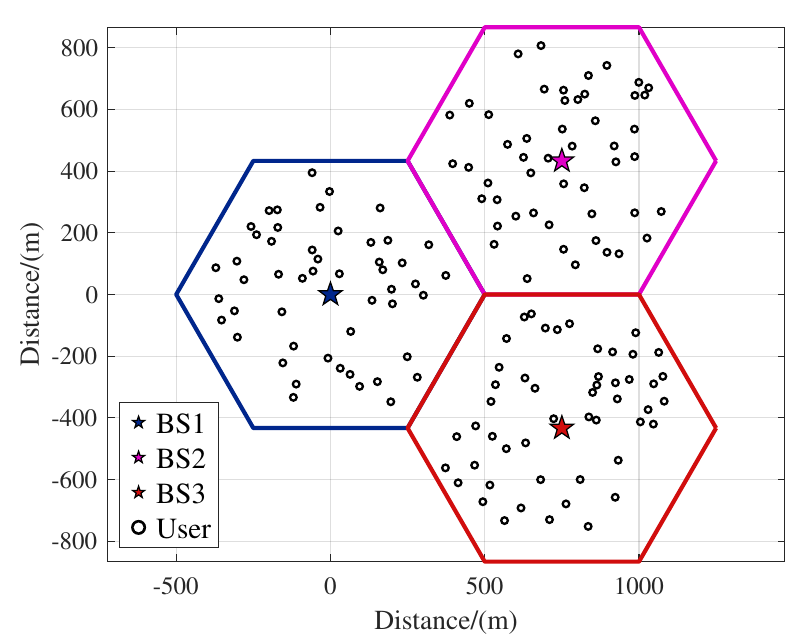}
        \caption{\small Scenario 1: {random user distribution}.}
        \label{fig3}
    \end{subfigure}
    \hfill
    \begin{subfigure}[t]{0.31\textwidth}
        \centering
        \includegraphics[width=\linewidth]{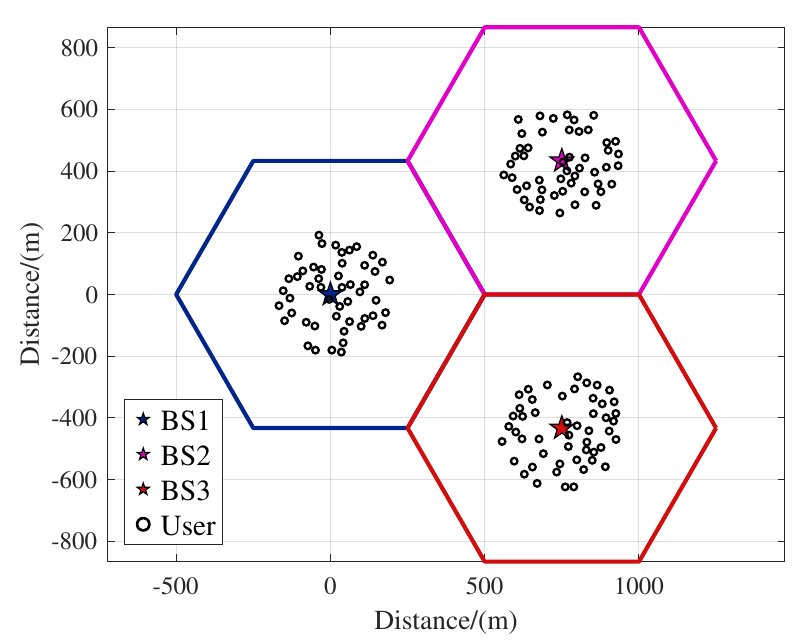}
        \caption{\small Scenario 2: {dense user distribution}.}
        \label{fig4}
    \end{subfigure}
    \hfill
    \begin{subfigure}[t]{0.36\textwidth}
        \centering
		\includegraphics[width=0.85\linewidth]{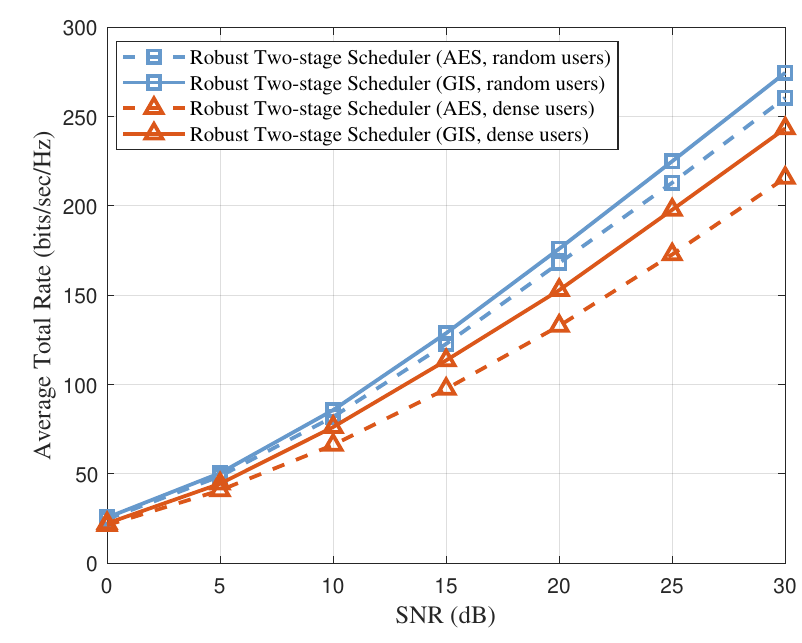}
		\caption{\small Comparison of sum-rate performance.}
		\label{fig5}
    \end{subfigure}

    \caption{The sum-rate performance for different active user selection algorithms under different user distributions.}
    \label{fig4444}
\end{figure*}

\begin{table}[t]
    \caption{Basic System Parameters}
    \centering
    \renewcommand{\arraystretch}{1}
    \begin{tabular}{c c} 
        \toprule
        \textbf{System Parameters} & \textbf{Values} \\
        \midrule
        Center frequency \(f_c\) & $6.7$ GHz \\
        Number of cells \(L\) & $3$ \\
        BS height \(H_{\text{tx}}\) & $25$ m \\
        User height \(H_{\text{rx}}\) & $1.5$ m \\
        Number of users per cell \(K\) & $50$ \\
        Number of scheduled users per cell \({\bar{K}}\) & $5/10/15$ \\
		Number of active users selected per cell \({{K'}}\) & $20$ \\
        Number of antennas at BS \(N\) & $32$ \\
        Number of sampling points per subregion \(S\) & $6/9/12/15 $\\
        \bottomrule
    \end{tabular}
    \label{table1}
\end{table}

	\section{Simulation Results}\label{5}	
	\subsection{Simulation Setup}\label{5-1}
	
	In this section, we present the simulation results to illustrate
	the performance of our proposed algorithm.
	Since
	QuaDRiGa can generate time-varying massive MIMO
	channels \cite{jaeckel2014quadriga} that meet third Generation Partnership Project (3GPP) New Radio (NR) specifications \cite{3gpp2018study} and have been validated by various field tests, 
        \color{black}
    we adopt
	QuaDRiGa for channel generation in the 3GPP urban
	macro (UMa) non-line-of-sight (NLOS) scenario in simulations, where the distance-dependent attenuation is modeled via the large-scale fading components, including path loss and log-normal shadow fading \cite{3gpp2018study}, so that both the desired and interfering links decay with increasing BS-UE distance.
    \color{black}
    We specifically select $6.7$ GHz as a representative frequency in the upper $6$ GHz band, which has been shown in our previous work \cite{hou2025tensor1} to achieve a favorable balance between bandwidth availability and propagation loss.
	Unless otherwise specified, the system parameters
	are selected according to 3GPP specifications,
	as shown in Table \ref{table1}. The signal-to-noise ratio (SNR) is defined as
	$\text{SNR} = 
	\frac{\displaystyle
		|\mathbf{w}^\text{H}\mathbf{h}|^2
	}{
		\displaystyle
		\|\mathbf{w}^\text{H}\|^2\sigma^2
	}$.

       \vspace{-5pt}

	\subsection{Benchmarks
	}\label{5-2}
	
	To evaluate the performance of the proposed algorithm, we select the following state-of-the-art approaches as benchmarks:
	
	\begin{itemize}
		\item \textbf{Greedy Scheduler\cite{dimic2005downlink}:} This algorithm  adopts a greedy approach 
		to maximize the sum rate of scheduled users  across all cells based on the  ICSI.

		\item \textbf{Random Scheduler\cite{liu2015low}:} This algorithm  randomly selects users without considering ICI.
		
		\item \textbf{SUS Scheduler\cite{yoo2006optimality}:} This algorithm   maintains semi-orthogonality among the channels of scheduled users within each cell based on the ICSI while ignoring ICI.

	\end{itemize}
	
	To better evaluate the performance of these algorithms, we apply the above user scheduling algorithms and compare the results against our proposed algorithm. 
	In the evaluation process, we take the system sum-rate as the key performance indicator and adopt the MMSE receiver \cite{price2007communication}.

	\subsection{Simulation Results}


	\subsubsection{Visualization of Grid-based US-CKM}

To facilitate more effective user scheduling, the coverage area of each BS is discretized into grids, and a grid-based US-CKM is constructed. We extract the statistical channel gain and the statistical channel correlation to jointly characterize link quality and spatial correlations among users.
Fig. \ref{fig3333} presents the visualization of these grid-based US-CKM. 
Specifically, Fig. \ref{fig7.1} illustrates the three-cell layout, the positions of the associated BSs, and the grid partitioning of the coverage area.
Fig. \ref{fig7.2} shows the statistical channel gain map for all grids, reflecting the link quality at different geographic coordinates. Fig. \ref{fig7.3} depicts the statistical channel correlation  between grids, revealing the degree of spatial channel similarity between users.
For grids located at the cell boundaries, we follow the conventional CKM treatment \cite{wu2023environment,xu2024much}  by assigning each grid to the nearest BS. This ensures a unique BS association for all grids, while the impact of ICI is inherently captured through the statistical channel gains and correlations.

 	\subsubsection{The Sum-rate Performance in Different User Distribution Scenarios}

	To comprehensively evaluate the performance of the two intra-cell active user selection algorithms under varying interference levels, we select two typical user distribution patterns for simulations—{random user distribution} and {dense user distribution},  shown in Fig. \ref{fig3} and Fig. \ref{fig4}, respectively. 
The sum-rate
performance of the two active user selection algorithms in
different user distribution scenarios is shown in Fig. \ref{fig5}.
	It can be observed  that when users are widely dispersed, the spatial separation between scheduled and interfering users is large, facilitating effective spatial multiplexing and interference suppression. In contrast, dense user distributions severely limit  spatial multiplexing capabilities  and leads  to increased  inter‐user interference.

    \color{black}
	In the {random user distribution} scenario,
    \color{black}
     the sum rate of the GIS algorithm with $10$ scheduled users per cell is improved by $4$\% compared with the AES algorithm at an
	SNR of $30$ dB, indicating a marginal weakness of the AES algorithm in terms of sum-rate.
	In contrast, under the {dense user distribution} scenario, the sum-rate of the GIS algorithm with $10$ scheduled users per cell is $10$\% higher than that of the AES algorithm at an
	SNR of $30$ dB, highlighting the superior interference suppression and throughput enhancement of the GIS algorithm in such a strong interference environment.
	Since the GIS algorithm recomputes  cumulative correlation metrics  to account for the global interference of each excluded user, it significantly improves  the sum-rate performance, albeit  at the cost of a slightly increased computational complexity. To effectively adapt to varying interference conditions, we adopt a flexible user selection strategy: the AES algorithm is employed in low-interference scenarios to ensure efficiency, while the GIS algorithm is activated under high-interference conditions to enhance overall system throughput.

\begin{figure*}[t] 
    \centering
    \begin{subfigure}[t]{0.32\textwidth}
        \centering
  \includegraphics[width=\linewidth]{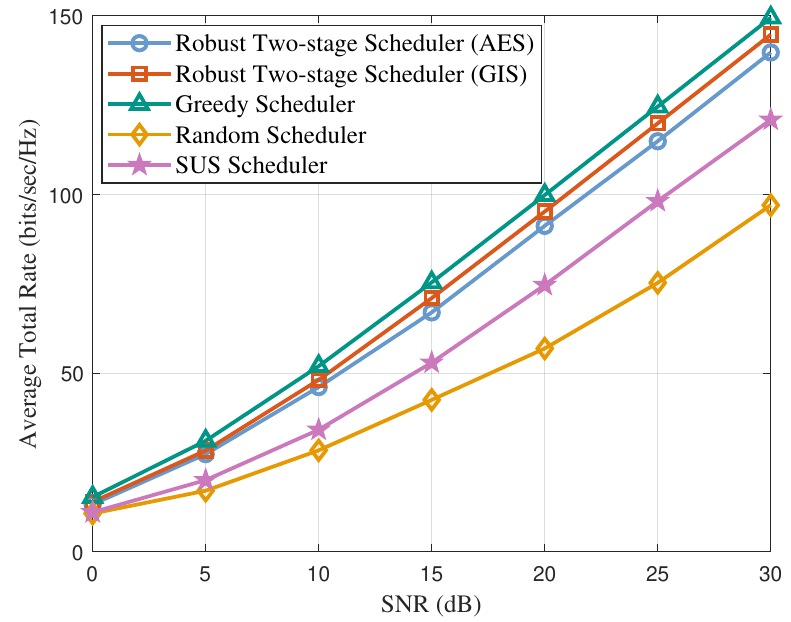}
        \caption{\small The scheduled users per cell are 5.}
        \label{fig6}
    \end{subfigure}
    \hfill
    \begin{subfigure}[t]{0.32\textwidth}
        \centering
\includegraphics[width=\linewidth]{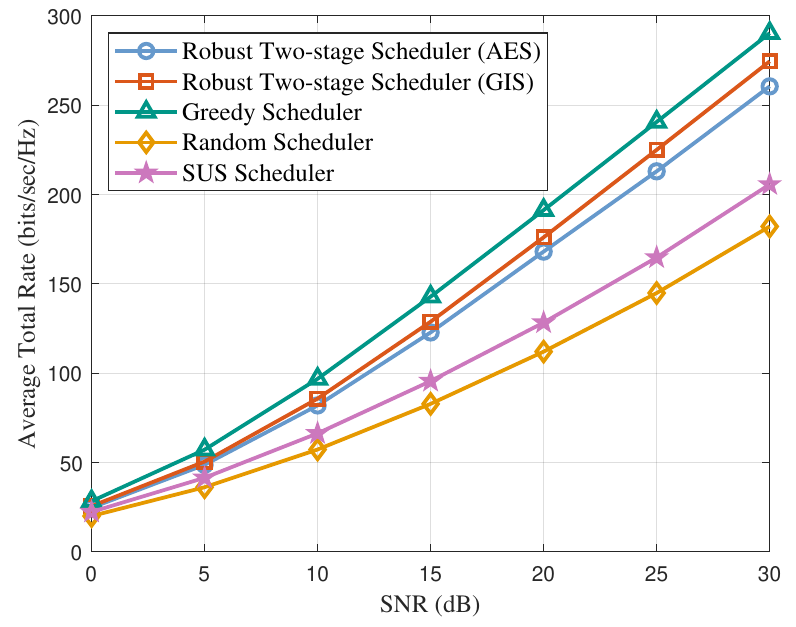}
        \caption{\small The scheduled users per cell are 10.}
        \label{fig7}
    \end{subfigure}
    \hfill
    \begin{subfigure}[t]{0.32\textwidth}
        \centering
\includegraphics[width=\linewidth]{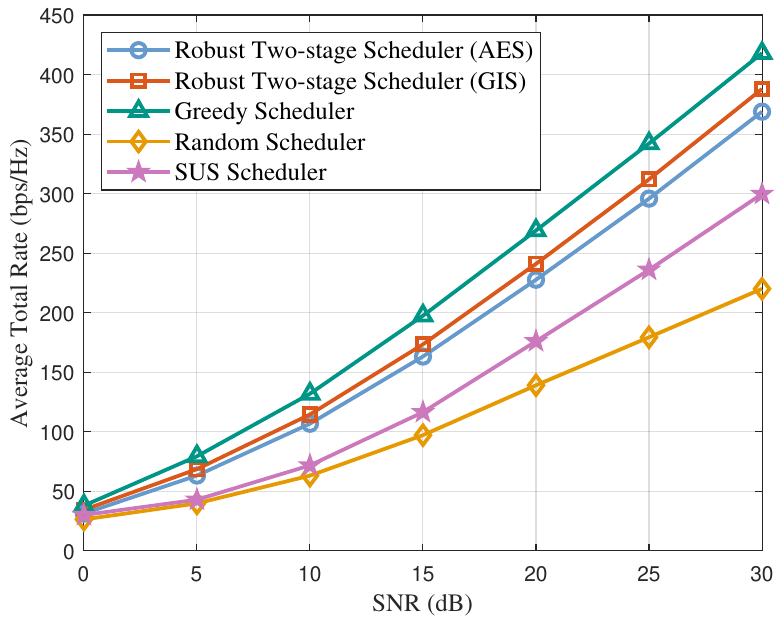}
        \caption{\small The scheduled users per cell are  15.}
        \label{fig6.1}
    \end{subfigure}
    \caption{The sum-rate performance for different algorithms when the number of scheduled users in each cell is different.}
    \label{fig5555}
\end{figure*}

\begin{figure*}[t] 
    \centering
    \setcounter{figure}{7}
    \begin{subfigure}[t]{0.32\textwidth}
        \centering
        \includegraphics[width=\linewidth]{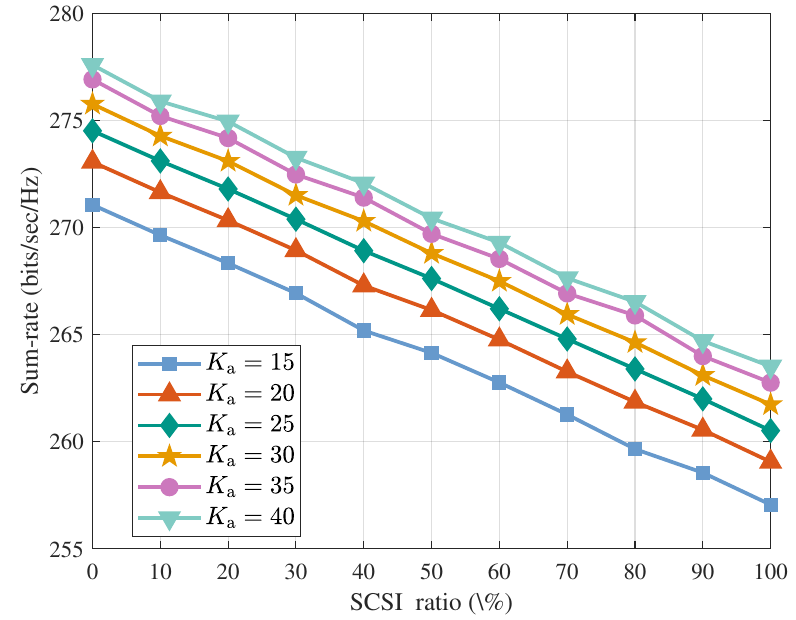}
        \caption{\small Sum-rate vs  
   SCSI proportions $\eta$ and ${K'}$.}
        \label{SCSI2}
    \end{subfigure}
    \hfill
    \begin{subfigure}[t]{0.32\textwidth}
        \centering
        \includegraphics[width=\linewidth]{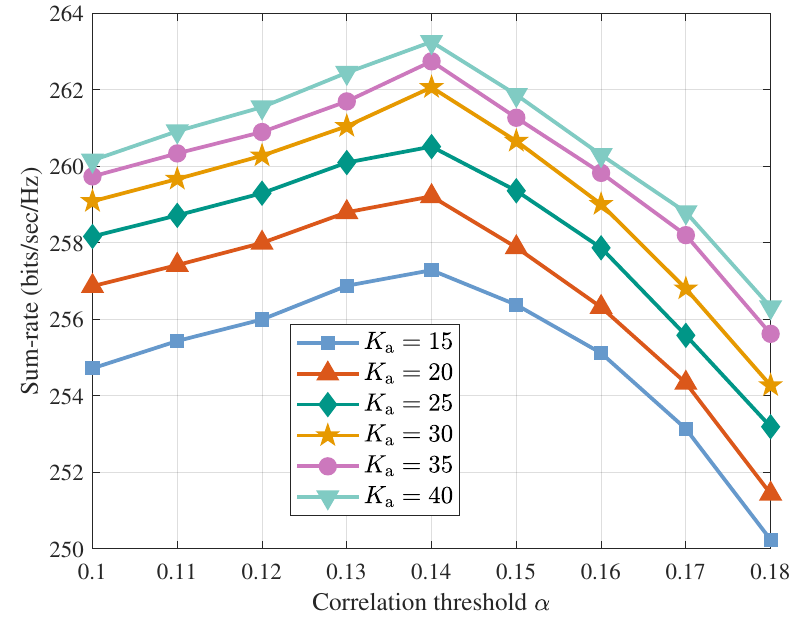}
        \caption{\small Sum-rate vs  ${\alpha}$ and ${K'}$.}
        \label{AES}
    \end{subfigure}
    \hfill
    \begin{subfigure}[t]{0.32\textwidth}
        \centering
        \includegraphics[width=\linewidth]{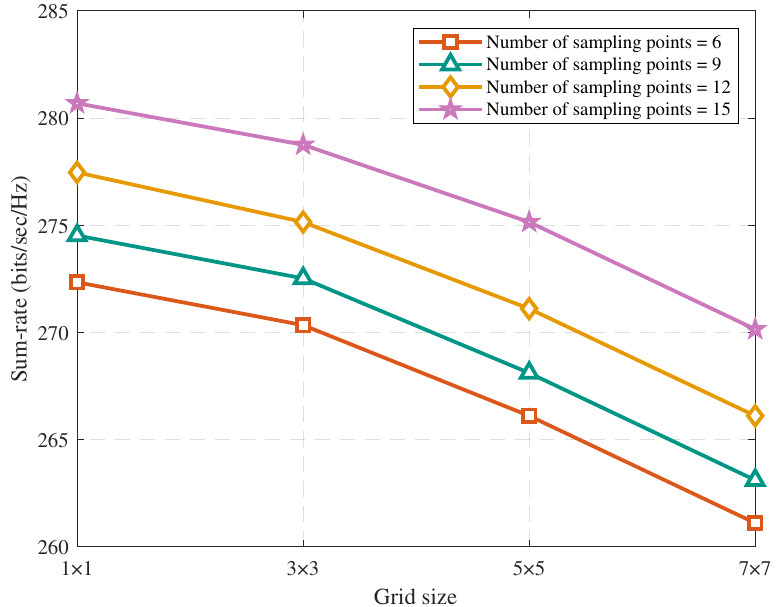}
        \caption{\small Sum-rate  vs  grid size and sampling density.}
        \label{GRID}
    \end{subfigure}
    \hfill
    \caption{The sum-rate performance with different ${\alpha}$, ${K'}$, SCSI proportions, grid size and sampling density.}
    \label{fig:combined}
\end{figure*}

 	\subsubsection{The Sum-rate Performance Comparison}

	Fig. \ref{fig6} shows the sum-rate performance of the proposed algorithm against various baselines across different SNRs. The \textbf{Greedy Scheduler}, leveraging ICSI and a greedy strategy to maximize system sum-rate, achieves the best result of 149.4 bits/sec/Hz. In contrast, the \textbf{Random Scheduler} and \textbf{SUS Scheduler}, both neglecting ICI in multi-cell systems, suffer notable losses of 35.1\% and 19.1\%, respectively. Notably, the proposed Robust Two-Stage Scheduler with GIS or AES as the first-stage selection is close to the \textbf{Greedy Scheduler}, with only 3.15\% and 5.16\% degradation, demonstrating its superiority over existing approaches.

	To examine the performance robustness of the proposed algorithms under different user loads, we change the number of scheduled users per cell from
	$5$ to $10$ and $15$, with the results shown  in Fig. \ref{fig7} and Fig. \ref{fig6.1}.
	As the number of scheduled users increases, the sum-rate gap between our proposed algorithms and the \textbf{Greedy Scheduler} remains nearly constant, whereas the gap  relative to the \textbf{Random Scheduler} and the \textbf{SUS Scheduler} progressively widens, from approximately $35$\% and $19$\% to above $40$\% and $25$\%, respectively. 
	This phenomenon indicates that our proposed algorithms can more effectively utilize spectral resources and suppress multi-user interference in  the multi-cell systems, thus achieving superior spectral efficiency.

 	\subsubsection{The Sum-rate Performance vs Different Parameters}

To facilitate a clearer comparison between the proposed \textbf{Robust Two-stage Scheduler} and the conventional \textbf{Two-stage Scheduler} relying solely on SCSI, the sum-rate performances of both algorithms, together with the baseline schemes, under varying SNR conditions are illustrated in Fig.~\ref{SCSI1}.
	The \textbf{Two-Stage Scheduler} achieves a sum-rate of $260.5$ bits/sec/Hz at $30$ dB. After updating the unreliable SCSI, the \textbf{Robust Two-stage Scheduler} attains an improved sum-rate of $274.5$ bits/sec/Hz. 
	The proposed algorithms demonstrate a substantial superiority in sum-rate performance over both the \textbf{Random Scheduler} and the \textbf{SUS Scheduler}.

\begin{figure}[t] 
\centering 
\setcounter{figure}{6}
\includegraphics[width=0.9\linewidth]{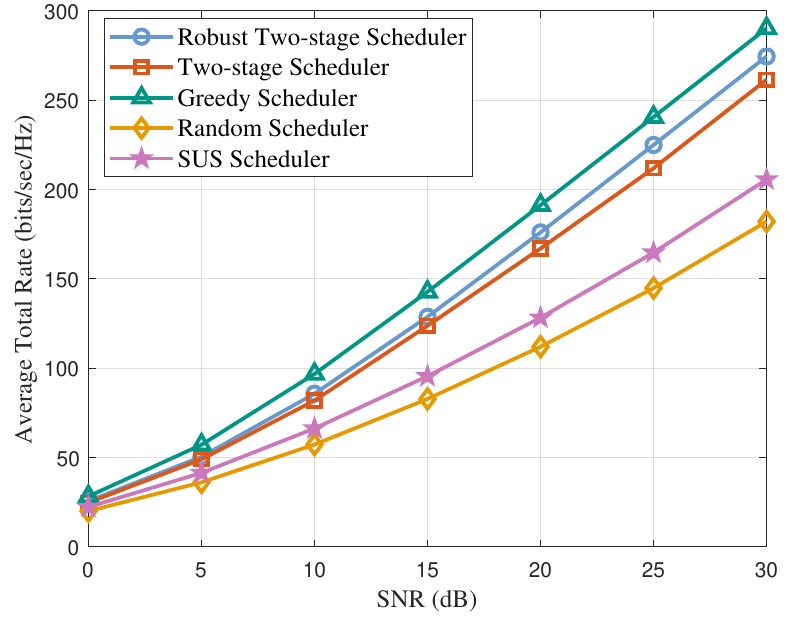} \caption{\small The sum-rate performance for two-stage algorithm before and after the update of unreliable SCSI.} 
\label{SCSI1} 
\end{figure}

	By adjusting the correlation variance threshold $\delta$, we can control the proportion of SCSI in the US-CKM. The impact of different SCSI proportions on the sum-rate performance at $30$ dB with $10$ users scheduled per cell is shown in Fig. \ref{SCSI2}. It is observed that an excessively high proportion of SCSI introduces a large amount of unreliable SCSI, thereby degrading the overall sum-rate performance of the proposed algorithm. Specifically, when the SCSI proportion reaches $100\%$, the sum-rate performance is reduced by approximately $14$ bits/sec/Hz compared to the case where only ICSI is used, highlighting the robust design in Section \ref{4}. When the SCSI proportions are set to $60$\% and $80$\%, the corresponding sum-rate performances are $266.2$ bits/sec/Hz and $263.4$ bits/sec/Hz, respectively—only slightly lower than the performance achieved with full ICSI. These results demonstrate that the proposed algorithm can effectively leverage a small amount of ICSI combined with a large portion of SCSI to maintain robust sum-rate performance under complex propagation conditions, while simultaneously achieving substantial reductions in computational complexity and CSI acquisition overhead.

Fig. \ref{AES} illustrates the impact of the hyperparameter correlation threshold ${\alpha}$ and the number of pre-selected candidate users ${K'}$ on the sum-rate performance of the AES algorithm.  It is evident that for any given ${K'}$ value, the sum rate performance shows a trend of first increasing and then decreasing with the increase in ${\alpha}$. This illustrates that the selection of ${\alpha}$ is a critical trade-off: an overly small value provides an insufficient number of candidates, while an excessively large value leads to severe intra-cell  interference.
	In addition, for a fixed value of ${\alpha}$, while increasing the number of pre-selected candidate users ${K'}$, from $25$ to $40$, the sum-rate only increases marginally from approximately $1.5$\%, indicating  the performance difference in sum-rate  is negligible. Therefore, we can confirm that our first-stage active user selection algorithms are capable of effectively reducing the candidate user set while incurring only a marginal degradation in sum-rate performance.

Fig. \ref{GRID} illustrates the impact of the hyperparameter, grid size and sampling density, on the sum-rate performance. 
For example, increasing the sampling points from $6$ to $12$ under a fixed grid size will result in an improvement of approximately 2.5\%,
indicating that sampling density is a key factor: more samples yield more reliable statistical channel estimates and, in turn, more accurate user scheduling.
Moreover, for a fixed sampling density, enlarging the grid from \(1\times 1\,\mathrm{m}\) to \(7\times 7\,\mathrm{m}\) reduces the sum rate by about \(3\%\), primarily due to the loss of spatial resolution in capturing fine-grained channel variations. These observations highlight that both parameters significantly affect system performance, and should be jointly optimized to strike a balance between communication efficiency and practical overhead.


\begin{table*}[t]
    \centering
    \caption{\sc\small Complexity Comparison}
    {
        \renewcommand{\arraystretch}{1.4}
        \begin{tabular}{l l l}
            \toprule
            Algorithm & Complexity & Multiplication Count \\
            \midrule
            Greedy Scheduler         & $\mathcal{O}\!\bigl(LK{\bar{K}}^{2}(N^{3}+KN^{2}+{\bar{K}}N^{2})\bigr)$ & $1.413\times10^{9}$ \\
            SUS Scheduler            & $\mathcal{O}\!\bigl(LK{\bar{K}}N\bigr)$                                     & $4.800\times10^{4}$ \\
            Random Scheduler         & $\mathcal{O}\!\bigl(1\bigr)$                                             & $1.000\times10^{0}$ \\
            Two-stage Scheduler (AES) & $\mathcal{O}\!\bigl(L K{K^{\prime}}^{2}+L^2{\bar{K}}^2K^{\prime}+L^2{\bar{K}}^3\bigr)$ & $8.700\times10^{4}$ \\
            Two-stage Scheduler (GIS) & $\mathcal{O}\!\bigl(LK^3+L^2{\bar{K}}^2K^{\prime}+L^2{\bar{K}}^3\bigr)$             & $4.020\times10^{5}$\\
            Robust Scheduler (AES) & $\mathcal{O}\!\bigl(L^2 (1-\eta)K+L^3(1-\eta)^2K^2 + L K{K^{\prime}}^{2}+L^2{\bar{K}}^2K^{\prime}+L^2{\bar{K}}^3\bigr)$ & $1.116\times10^{5}$ \\
            Robust  Scheduler (GIS) & $\mathcal{O}\!\bigl(L^2 (1-\eta)K+L^3(1-\eta)^2K^2+LK^3+L^2{\bar{K}}^2K^{\prime}+L^2{\bar{K}}^3\bigr)$ & $4.266\times10^{5}$ \\
            \bottomrule
        \end{tabular}
    }
    \label{table2}
\end{table*}

\begin{table}[t]
	\caption{CSI acquisition overhead.}
	\label{tab:csi_overhead1}
	\centering
	\renewcommand{\arraystretch}{1.15}
	\begin{tabular}{lll}
		\hline
		\textbf{Algorithm} & \textbf{CSI acquisition overhead} & \textbf{Count} \\
		\hline
		Two-stage Scheduler & $\mathcal{O}\!\left(0\right)$ & $0$ \\
		Robust  Scheduler & $\mathcal{O}\!\left(L^2 (1-\eta) K\right)$ & $135$ \\
		Greedy Scheduler & $\mathcal{O}\!\left(L^2 K\right)$ & $450$ \\
		SUS Scheduler & $\mathcal{O}\!\left(L K\right)$ & $150$ \\
		\hline
	\end{tabular}
\end{table}

\begin{table}[t]
	\caption{Information exchange overhead.}
	\label{tab:info_exchange_overhead_overhead2}
	\centering
	\renewcommand{\arraystretch}{1.15}
	\begin{tabular}{lll}
		\hline
		\textbf{Algorithm} & \textbf{Information exchange overhead} & \textbf{Count} \\
		\hline
		Two-stage Scheduler & $\mathcal{O}\!\left(LK'\right)$ & $60$ \\
		Robust  Scheduler & $\mathcal{O}\!\left((2-\eta)LK' + L^3(1-\eta)K'\right)$ & $240$ \\
		Greedy Scheduler & $\mathcal{O}\!\left(L^2KN\right)$ & $14400$ \\
		SUS Scheduler & $\mathcal{O}\!\left(0\right)$ & $0$ \\
		\hline
	\end{tabular}
\end{table}

    \vspace{-5pt}

	\subsection{Computational Complexity and Signaling Overhead}\label{5-4}

\subsubsection{Computational Complexity}\label{5-4-1}

	To  assess the computational overhead of the different algorithms, the specific analysis of complexity is shown as follows.
	The proposed \textbf{Two-Stage Scheduler} operates in two stages: when the AES is used in the first stage, the complexity is $\mathcal{O}\bigl(LK{K^{\prime}}^{2}\bigr)$, whereas employing the GIS method yields a complexity of $\mathcal{O}\bigl(LK^3\bigr)$. 
	In contrast to the AES, the complexity of GIS is dominated by the iterative user elimination process in Algorithm \ref{alg:GIS}, which repeatedly removes users from the full set until the number of remaining users is reduced to $K^{\prime}$ per cell.
	The ICCS executed in the second stage adds a further complexity of $O(L^2{\bar{K}}^2K^{\prime}+L^2{\bar{K}}^3)$, which is primarily determined by the computation in \eqref{eq:17}.
	Compared with the \textbf{Two-Stage Scheduler}, the \textbf{Robust Two-stage Scheduler} introduces an additional CSI-update step, which incurs an extra computational cost of $\mathcal{O}(L^2 (1-\eta)K+L^3(1-\eta)^2K^2)$ for updating channel gain and channel correlation metrics.
	The complexity of the \textbf{Greedy Scheduler} is  $\mathcal{O}\!\bigl(LK{\bar{K}}^{2}(N^{3}+KN^{2}+{\bar{K}}N^{2})\bigr)$, while
	the complexity of the \textbf{SUS Scheduler} and the \textbf{Random Scheduler} are $\mathcal{O}\!\bigl(LK{\bar{K}}N\bigr)$ and $\mathcal{O}\!\bigl(1)$, respectively.
	In this analysis, $K$ represents the number of users per cell, ${\bar{K}}$ denotes the maximum number of scheduled users per cell, $N$ is the number of transmit antennas, $\eta$ is the proportion of SCSI in all CSI, and ${K'}$ represents the number of users selected during the active user selection process. 
	The computational complexities of several algorithms are provided in Table~\ref{table2}.

	It is noted that  the \textbf{Greedy Scheduler} exhibits high computational complexity, rendering it difficult to apply in practical scenarios. The \textbf{SUS scheduler} ensures only channel orthogonality between new and selected users to reduce computational overhead, but this neglect of ICI leads to a degradation in sum-rate performance.
	By comparison, our proposed \textbf{Robust Two-stage Scheduler} achieves sum-rate performance comparable to the conventional \textbf{Greedy Scheduler}, but with significantly reduced computational complexity.

  \color{black}  

\subsubsection{CSI Acquisition Overhead}\label{5-4-2}

Table~\ref{tab:csi_overhead1} compares the CSI acquisition overhead of different scheduling algorithms. 
Specifically, 
for the proposed two-stage algorithm, the online scheduling process relies only on user location information and the offline-constructed CKM database, and therefore incurs almost no additional CSI acquisition overhead. For the proposed robust algorithm, real-time CSI updates are triggered for users located in unreliable grids, resulting in a CSI acquisition overhead of $\mathcal{O}\!\left(L^2(1-\eta)K\right)$. For the conventional greedy algorithm, instantaneous global CSI of all users must be acquired at all BSs, and thus its CSI acquisition overhead reaches $\mathcal{O}\!\left(L^2K\right)$. For the conventional SUS algorithm, only the CSI from each user to its local serving BS is required, and hence the corresponding CSI acquisition overhead is $\mathcal{O}\!\left(LK\right)$.

Overall, the proposed two-stage scheduler incurs no online CSI acquisition overhead, while the robust two-stage scheduler can significantly reduce the CSI acquisition burden while achieving sum-rate performance comparable to that of the greedy scheduler.

\color{black}

\subsubsection{Information  Exchange Overhead}\label{5-4-3}

\color{black}

Table~\ref{tab:info_exchange_overhead_overhead2} compares the information exchange overhead of different scheduling algorithms. 
Specifically, for the proposed two-stage algorithm, each local BS only needs to report the location information of the $LK'$ candidate users to the CU for centralized scheduling, and thus its information exchange overhead is $\mathcal{O}\!\left(LK'\right)$. For the proposed robust algorithm, in addition to the location information reporting, the instantaneous channel gains and channel correlations of users located in unreliable grids also need to be uploaded, resulting in a total information exchange overhead of $\mathcal{O}\!\left((1-\eta)LK' + L^3(1-\eta)K' + LK'\right)$. For the conventional centralized greedy scheduling algorithm, each local BS is required to upload the global instantaneous CSI of all users to the CU for centralized scheduling, and therefore its information exchange overhead is $\mathcal{O}\!\left(L^2KN\right)$. For the SUS scheduling algorithm, user scheduling is performed locally at each BS without any additional information exchange, and hence its information exchange overhead is negligible.

Overall, the proposed two-stage scheduler incurs only low information exchange overhead, while the proposed robust two-stage scheduler can achieve an aggregate-rate performance comparable to that of the greedy scheduler while substantially reducing the information exchange overhead. Although the SUS algorithm requires no additional information exchange overhead, its sum-rate performance is 19.1\% lower than that of the proposed robust two-stage scheduler in Fig.~\ref{fig6}, indicating that it is not well suited.

\color{black}

    \vspace{-5pt}

	\section{Conclusion}\label{6}
In this paper, we investigated the coordinated user
	scheduling  problem for multi-cell massive MIMO.
	Specifically,  we formulated the coordinated user scheduling problem with the spectral
	efficiency maximization criterion.
	By exploiting spatial consistency in wireless channels, we constructed a US-CKM to enable the location-based retrieval of the required SCSI, thereby avoiding real-time CSI acquisition.
	Building on the US-CKM framework, we propose a low-complexity two-stage coordinated user scheduling algorithm that mitigates inter-user interference in two steps.
	In the first stage,  active user selection shrinks  the set of candidate users, and  also effectively  suppresses intra-cell interference. In the second stage, inter-cell coordinated scheduling is performed to  effectively suppress ICI.
	To enhance robustness against imperfect SCSI in environments with dynamic scatterers, we proposed a  robustness-enhanced two-stage coordinated user scheduling algorithm.
	Numerical results illustrated that the
	proposed  algorithms achieve sum-rate
	performance comparable to the conventional greedy algorithm
	while significantly reducing computational complexity and
	signaling overhead.

\appendices
\section{Derivation of Equations (15)--(17)}\label{APP}

\color{black}

Starting from \eqref{eq:14}, we provide a detailed derivation of \eqref{eq:15} and \eqref{eq:16}, which follows the standard projection-based orthogonalization procedure widely adopted in SUS-inspired user selection and scheduling algorithms \cite{yoo2006optimality,benmimoune2015joint}.
Building on this, the orthogonal residual channel is given by
\begin{equation}
{\mathbf{h}}_{l,m,k}^{\perp}
= {\mathbf{h}}_{l,m,k}
- \sum_{q=1}^{L}\sum_{j=1}^{i-1}
{\rho}_{l,m,k,q,j}\,
\bigl\|{\mathbf{h}}_{l,m,k}\bigr\|\,
{\hat{\mathbf{h}}}_{l,q,j},
\label{eq:app_eq14}
\end{equation}
where $\hat{\mathbf{h}}_{l,q,j}=\mathbf{h}_{l,q,j}/\|\mathbf{h}_{l,q,j}\|$ denotes the unit-norm channel direction.
Since $\hat{\mathbf{h}}_{l,m,k}=\mathbf{h}_{l,m,k}/\|\mathbf{h}_{l,m,k}\|$, we have
$\mathbf{h}_{l,m,k}=\|\mathbf{h}_{l,m,k}\|\,\hat{\mathbf{h}}_{l,m,k}$.
Substituting this identity into \eqref{eq:app_eq14} yields
\begin{align}
\mathbf{h}_{l,m,k}^{\perp}
&=
\|\mathbf{h}_{l,m,k}\|\hat{\mathbf{h}}_{l,m,k}
-\|\mathbf{h}_{l,m,k}\|
\sum_{q=1}^{L}\sum_{j=1}^{i-1}
\rho_{l,m,k,q,j}\hat{\mathbf{h}}_{l,q,j} \nonumber\\
&=
\|\mathbf{h}_{l,m,k}\|
\left(
\hat{\mathbf{h}}_{l,m,k}
-\sum_{q=1}^{L}\sum_{j=1}^{i-1}
\rho_{l,m,k,q,j}\hat{\mathbf{h}}_{l,q,j}
\right).
\label{eq:app_factor}
\end{align}
Taking the Euclidean norm on both sides of \eqref{eq:app_factor} immediately leads to
\begin{equation}
\|{\mathbf{h}}_{l,m,k}^{\perp}\|
=
\|{\mathbf{h}}_{l,m,k}\|\,
\Bigl\|
\hat{\mathbf{h}}_{l,m,k}
-\sum_{q=1}^{L}\sum_{j=1}^{i-1}
{\rho}_{l,m,k,q,j}\,\hat{\mathbf{h}}_{l,q,j}
\Bigr\|,
\label{eq:app_eq15}
\end{equation}
which coincides with \eqref{eq:15}.
Define the residual direction vector as
\begin{equation}
\mathbf{r}_{l,m,k}
=
\hat{\mathbf{h}}_{l,m,k}
-\sum_{q=1}^{L}\sum_{j=1}^{i-1}
{\rho}_{l,m,k,q,j}\,\hat{\mathbf{h}}_{l,q,j}.
\label{eq:app_residual_def}
\end{equation}
Then, its squared norm can be expanded as
\begin{align}
&\|\mathbf{r}_{l,m,k}\|^{2}
=
\left(
\hat{\mathbf{h}}_{l,m,k}
-\sum_{q=1}^{L}\sum_{j=1}^{i-1}
\rho_{l,m,k,q,j}\hat{\mathbf{h}}_{l,q,j}
\right)^{\!\mathrm H}
\nonumber\\[-2pt]
&\quad\times
\left(
\hat{\mathbf{h}}_{l,m,k}
-\sum_{q'=1}^{L}\sum_{j'=1}^{i-1}
\rho_{l,m,k,q',j'}\hat{\mathbf{h}}_{l,q',j'}
\right)
\nonumber\\
&=
\underbrace{\hat{\mathbf{h}}_{l,m,k}^{\mathrm H}\hat{\mathbf{h}}_{l,m,k}}_{=\,1}
-2\,\Re\!\left\{
\sum_{q=1}^{L}\sum_{j=1}^{i-1}
\rho_{l,m,k,q,j}\,
\hat{\mathbf{h}}_{l,q,j}^{\mathrm H}\hat{\mathbf{h}}_{l,m,k}
\right\}
\nonumber\\
&\quad+
\sum_{q=1}^{L}\sum_{j=1}^{i-1}
\sum_{q'=1}^{L}\sum_{j'=1}^{i-1}
\rho_{l,m,k,q,j}\rho_{l,m,k,q',j'}\,
\hat{\mathbf{h}}_{l,q,j}^{\mathrm H}\hat{\mathbf{h}}_{l,q',j'}.
\label{eq:app_expand}
\end{align}

In the proposed two-stage framework, the AES/GIS procedure ensures that the previously selected
unit-norm channel directions
$\{\hat{\mathbf{h}}_{l,q,j}\}_{q=1,\dots,L}^{j=1,\dots,i-1}$
are weakly correlated and can be regarded as approximately orthonormal, i.e.,
\begin{equation}
\hat{\mathbf{h}}_{l,q,j}^{\mathrm H}\hat{\mathbf{h}}_{l,q',j'}
\approx
\begin{cases}
1, & (q,j)=(q',j'),\\
0, & (q,j)\neq(q',j').
\end{cases}
\label{eq:approx_orthonormal}
\end{equation}
As a consequence, the double summation term in \eqref{eq:app_expand} can be approximated by discarding
the cross-terms, yielding
\begin{equation}
\begin{aligned}
&\sum_{q=1}^{L}\sum_{j=1}^{i-1}
\sum_{q'=1}^{L}\sum_{j'=1}^{i-1}
\rho_{l,m,k,q,j}\rho_{l,m,k,q',j'}\,
\hat{\mathbf{h}}_{l,q,j}^{\mathrm H}\hat{\mathbf{h}}_{l,q',j'} \\
&\qquad\approx
\sum_{q=1}^{L}\sum_{j=1}^{i-1}
\left|\rho_{l,m,k,q,j}\right|^{2}.
\end{aligned}
\label{eq:app_term3_simplify}
\end{equation}
This approximation is justified by the fact that the estimated channel vectors associated with different index pairs $(q,j)\neq(q',j')$ are typically weakly correlated in the considered multi-user multi-cell setting; hence, the corresponding off-diagonal inner products concentrate around zero in an average sense, making their aggregate contribution negligible compared with the diagonal energy terms.
Moreover, recalling the normalized channel correlation definition
\begin{equation}
\rho_{l,m,k,q,j}
=
\frac{\left|\mathbf{h}_{l,m,k}^{\mathrm H}\mathbf{h}_{l,q,j}\right|}
{\left\|\mathbf{h}_{l,m,k}\right\|\left\|\mathbf{h}_{l,q,j}\right\|}
=
\left|\hat{\mathbf{h}}_{l,q,j}^{\mathrm H}\hat{\mathbf{h}}_{l,m,k}\right|,
\label{eq:app_rho_abs}
\end{equation}
the second term in \eqref{eq:app_expand} can be approximated as
\begin{equation}
\begin{aligned}
-2\,\Re\!\left\{
\sum_{q=1}^{L}\sum_{j=1}^{i-1}
\rho_{l,m,k,q,j}\,
\hat{\mathbf{h}}_{l,q,j}^{\mathrm H}\hat{\mathbf{h}}_{l,m,k}
\right\}
&\approx \\
-2\sum_{q=1}^{L}\sum_{j=1}^{i-1}
\left|\rho_{l,m,k,q,j}\right|^{2}.
\end{aligned}
\label{eq:app_term2_simplify}
\end{equation}

Substituting \eqref{eq:app_term3_simplify} into the last term of \eqref{eq:app_expand} and applying
\eqref{eq:app_term2_simplify} to the real-part term, we obtain
\begin{align}
\|\mathbf{r}_{l,m,k}\|^{2}
&\approx
\underbrace{\hat{\mathbf{h}}_{l,m,k}^{\mathrm H}\hat{\mathbf{h}}_{l,m,k}}_{=\,1}
-2\sum_{q=1}^{L}\sum_{j=1}^{i-1}|\rho_{l,m,k,q,j}|^{2}
\nonumber\\
&\quad+
\sum_{q=1}^{L}\sum_{j=1}^{i-1}|\rho_{l,m,k,q,j}|^{2}
\nonumber\\
&=
1-\sum_{q=1}^{L}\sum_{j=1}^{i-1}|\rho_{l,m,k,q,j}|^{2}.
\label{eq:app_eq16_final}
\end{align}
Therefore, the squared norm of the residual direction is approximated by the complement of the total
projection energy onto the previously selected (approximately orthonormal) unit-norm channel directions.
Finally, recalling the definition of $\mathbf{r}_{l,m,k}$ in \eqref{eq:app_residual_def}, the above result
directly yields \eqref{eq:16}.

\color{black}

\vspace{-5pt}

\bibliographystyle{IEEEtran}
\bibliography{ref}

\begin{thebibliography}{10}
\providecommand{\url}[1]{#1}
\csname url@samestyle\endcsname
\providecommand{\newblock}{\relax}
\providecommand{\bibinfo}[2]{#2}
\providecommand{\BIBentrySTDinterwordspacing}{\spaceskip=0pt\relax}
\providecommand{\BIBentryALTinterwordstretchfactor}{4}
\providecommand{\BIBentryALTinterwordspacing}{\spaceskip=\fontdimen2\font plus
\BIBentryALTinterwordstretchfactor\fontdimen3\font minus
  \fontdimen4\font\relax}
\providecommand{\BIBforeignlanguage}[2]{{%
\expandafter\ifx\csname l@#1\endcsname\relax
\typeout{** WARNING: IEEEtran.bst: No hyphenation pattern has been}%
\typeout{** loaded for the language `#1'. Using the pattern for}%
\typeout{** the default language instead.}%
\else
\language=\csname l@#1\endcsname
\fi
#2}}
\providecommand{\BIBdecl}{\relax}
\BIBdecl

\bibitem{xie2025multi}
S.~Xie and X.~Huang, ``Multi-cell cooperative transmission for {MU-NOMA}
  networks,'' \emph{Wirel. Netw.}, vol.~31, no.~3, pp. 2457--2475, JAN. 2025.

\bibitem{hou2024joint}
H.~Hou, Y.~Wang, X.~Yi, W.~Wang, and S.~Jin, ``Joint beam alignment and doppler
  estimation for fast time-varying wideband {mmWave} channels,'' \emph{IEEE
  Trans. Wireless Commun.}, vol.~23, no.~9, pp. 10\,895--10\,910, SEP. 2024.

\bibitem{hou2024tensor}
H.~Hou, Y.~Wang, Y.~Zhu, X.~Yi, W.~Wang, D.~Slock, and S.~Jin, ``A
  tensor-structured approach to dynamic channel prediction for massive {MIMO}
  systems with temporal non-stationarity,'' \emph{arXiv preprint
  arXiv:2412.06713}, 2024.

\bibitem{hamza2013survey}
A.~S. Hamza, S.~S. Khalifa, H.~S. Hamza, and K.~Elsayed, ``A survey on
  inter-cell interference coordination techniques in {OFDMA}-based cellular
  networks,'' \emph{IEEE Commun. Surveys Tuts.}, vol.~15, no.~4, pp.
  1642--1670, OCT. 2013.

\bibitem{zhang2011weighted}
H.~Zhang, L.~Venturino, N.~Prasad, P.~Li, S.~Rangarajan, and X.~Wang,
  ``Weighted sum-rate maximization in multi-cell networks via coordinated
  scheduling and discrete power control,'' \emph{IEEE J. Sel. Areas Commun.},
  vol.~29, no.~6, pp. 1214--1224, JUN. 2011.

\bibitem{sun2015beam}
C.~Sun, X.~Gao, S.~Jin, M.~Matthaiou, Z.~Ding, and C.~Xiao, ``Beam division
  multiple access transmission for massive {MIMO} communications,'' \emph{IEEE
  Trans. Commun.}, vol.~63, no.~6, pp. 2170--2184, JUN. 2015.

\bibitem{wang2024towards}
Y.~Wang, H.~Hou, X.~Yi, W.~Wang, and S.~Jin, ``Towards unified {AI} models for
  {MU-MIMO} communications: A tensor equivariance framework,'' \emph{IEEE
  Trans. Wireless Commun.}, JUL. 2025, early Access.

\bibitem{wu2023low}
S.~Wu, G.~Sun, Y.~Wang, L.~You, W.~Wang, and R.~Ding, ``Low-complexity user
  scheduling for {LEO} satellite communications,'' \emph{IET Commun.}, vol.~17,
  no.~12, pp. 1368--1383, APR. 2023.

\bibitem{hou2025tensor1}
H.~Hou, Y.~Wang, X.~Yi, W.~Wang, D.~Slock, and S.~Jin, ``Tensor-structured
  {Bayesian} channel prediction for upper mid-band {XL-MIMO} systems,''
  \emph{arXiv preprint arXiv:2508.08491}, 2025.

\bibitem{dimic2005downlink}
G.~Dimic and N.~D. Sidiropoulos, ``On downlink beamforming with greedy user
  selection: Performance analysis and a simple new algorithm,'' \emph{IEEE
  Trans. Signal Process.}, vol.~53, no.~10, pp. 3857--3868, OCT. 2005.

\bibitem{jiang2006greedy}
J.~Jiang, R.~M. Buehrer, and W.~H. Tranter, ``Greedy scheduling performance for
  a zero-forcing dirty-paper coded system,'' \emph{IEEE Trans. Commun.},
  vol.~54, no.~5, pp. 789--793, MAY 2006.

\bibitem{yoo2006optimality}
T.~Yoo and A.~Goldsmith, ``On the optimality of multiantenna broadcast
  scheduling using zero-forcing beamforming,'' \emph{IEEE J. Sel. Areas
  Commun.}, vol.~24, no.~3, pp. 528--541, MAR. 2006.

\bibitem{benmimoune2015joint}
M.~Benmimoune, E.~Driouch, W.~Ajib, and D.~Massicotte, ``Joint transmit antenna
  selection and user scheduling for massive {MIMO} systems,'' in \emph{Proc.
  IEEE Wireless Commun. Netw. Conf. (WCNC)}, 2015, pp. 381--386.

\bibitem{liu2015low}
H.~Liu, H.~Gao, S.~Yang, and T.~Lv, ``Low-complexity downlink user selection
  for massive {MIMO} systems,'' \emph{IEEE Syst. J.}, vol.~11, no.~2, pp.
  1072--1083, JUN. 2015.

\bibitem{huang2012decremental}
S.~Huang, H.~Yin, H.~Li, and V.~C. Leung, ``Decremental user selection for
  large-scale multi-user {MIMO} downlink with zero-forcing beamforming,''
  \emph{IEEE Wireless Commun. Lett.}, vol.~1, no.~5, pp. 480--483, OCT. 2012.

\bibitem{farsaei2019improved}
A.~Farsaei, A.~Alvarado, F.~M. Willems, and U.~Gustavsson, ``An improved
  dropping algorithm for line-of-sight massive {MIMO} with max-min power
  control,'' \emph{IEEE Commun. Lett.}, vol.~23, no.~6, pp. 1109--1112, JUN.
  2019.

\bibitem{li2014multi}
G.~Y. Li, J.~Niu, D.~Lee, J.~Fan, and Y.~Fu, ``Multi-cell coordinated
  scheduling and {MIMO} in {LTE},'' \emph{IEEE Commun. Surveys Tuts.}, vol.~16,
  no.~2, pp. 761--775, AUG. 2014.

\bibitem{kwan2010survey}
R.~Kwan and C.~Leung, ``A survey of scheduling and interference mitigation in
  {LTE},'' \emph{J. Electr. Comput. Eng.}, vol. 2010, no.~1, p. 273486, JUL.
  2010.

\bibitem{venturino2009coordinated}
L.~Venturino, N.~Prasad, and X.~Wang, ``Coordinated scheduling and power
  allocation in downlink multicell {OFDMA} networks,'' \emph{IEEE Trans. Veh.
  Technol.}, vol.~58, no.~6, pp. 2835--2848, JUN. 2009.

\bibitem{sawahashi2010coordinated}
M.~Sawahashi, Y.~Kishiyama, A.~Morimoto, D.~Nishikawa, and M.~Tanno,
  ``Coordinated multipoint transmission/reception techniques for {LTE}-advanced
  coordinated and distributed {MIMO},'' \emph{IEEE Wireless Commun.}, vol.~17,
  no.~3, pp. 26--34, JUL. 2010.

\bibitem{wang2024soft}
Y.~Wang, H.~Hou, W.~Wang, X.~Yi, and S.~Jin, ``Soft demodulator for
  symbol-level precoding in coded multiuser {MISO} systems,'' \emph{IEEE Trans.
  Wireless Commun.}, vol.~23, no.~10, pp. 14\,819--14\,835, 2024.

\bibitem{lahoud2016energy}
S.~Lahoud, K.~Khawam, S.~Martin, G.~Feng, Z.~Liang, and J.~Nasreddine,
  ``Energy-efficient joint scheduling and power control in multi-cell wireless
  networks,'' \emph{IEEE J. Sel. Areas Commun.}, vol.~34, no.~12, pp.
  3409--3426, SEP. 2016.

\bibitem{sang2004coordinated}
A.~Sang, X.~Wang, M.~Madihian, and R.~D. Gitlin, ``Coordinated load balancing,
  handoff/cell-site selection, and scheduling in multi-cell packet data
  systems,'' in \emph{Proc. ACM MobiCom}, FEB. 2004, pp. 302--314.

\bibitem{sun2013interference}
S.~Sun, Q.~Gao, Y.~Peng, Y.~Wang, and L.~Song, ``Interference management
  through {CoMP} in {3GPP} {LTE}-advanced networks,'' \emph{IEEE Wireless
  Commun.}, vol.~20, no.~1, pp. 59--66, JAN. 2013.

\bibitem{cai2015joint}
Y.~Cai, J.~Zheng, Y.~Wei, Y.~Xu, and A.~Anpalagan, ``A joint game-theoretic
  interference coordination approach in uplink multi-cell {OFDMA} networks,''
  \emph{Wirel. Pers. Commun.}, vol.~80, pp. 1203--1215, SEP. 2015.

\bibitem{shan2024resource}
L.~Shan, S.~Gao, Y.~Yu, F.~Zhang, Y.~Hu, Y.~Wang, and M.~Chen, ``Resource
  allocation for multi-cell multi-timeslot transmission: Centralized and
  distributed algorithms,'' \emph{IEEE Trans. Netw. Serv. Manag.}, vol.~21,
  no.~3, pp. 3021--3034, JUN. 2024.

\bibitem{cai2025cooperative}
Y.~Cai, Z.~Zhang, Y.~Huang, W.~Yu, X.~Nie, and H.~Liu, ``Cooperative resource
  allocation for {NOMA-MEC} multi-cell network,'' \emph{IEEE Trans. Veh.
  Technol.}, vol.~74, no.~6, pp. 9027--9042, FEB. 2025.

\bibitem{shi2020learning}
J.~Shi, W.~Wang, X.~Yi, J.~Wang, X.~Gao, Q.~Liu, and G.~Y. Li, ``Learning to
  compute ergodic rate for multi-cell scheduling in massive {MIMO},''
  \emph{IEEE Trans. Wireless Commun.}, vol.~20, no.~2, pp. 785--797, JUL. 2020.

\bibitem{ding2024improving}
X.~Ding, Y.~Ren, X.~Xie, Y.~Zou, M.~Jia, and G.~Zhang, ``Improving user
  capacity of satellite internet of things via joint user grouping and
  multi-beam processing,'' \emph{IEEE Trans. Wireless Commun.}, vol.~72, no.~7,
  pp. 3957--3969, APR. 2024.

\bibitem{zhang2017sum}
C.~Zhang, Y.~Huang, Y.~Jing, S.~Jin, and L.~Yang, ``Sum-rate analysis for
  massive {MIMO} downlink with joint statistical beamforming and user
  scheduling,'' \emph{IEEE Trans. Wireless Commun.}, vol.~16, no.~4, pp.
  2181--2194, SEP. 2017.

\bibitem{zeng2021toward}
Y.~Zeng and X.~Xu, ``Toward environment-aware {6G} communications via channel
  knowledge map,'' \emph{IEEE Wireless Commun.}, vol.~28, no.~3, pp. 84--91,
  JUN. 2021.

\bibitem{wu2023environment}
D.~Wu, Y.~Zeng, S.~Jin, and R.~Zhang, ``Environment-aware hybrid beamforming by
  leveraging channel knowledge map,'' \emph{IEEE Trans. Wireless Commun.},
  vol.~23, no.~5, pp. 4990--5005, OCT. 2023.

\bibitem{xu2024much}
X.~Xu and Y.~Zeng, ``How much data is needed for channel knowledge map
  construction?'' \emph{IEEE Trans. Wireless Commun.}, vol.~23, no.~10, pp.
  13\,011--13\,021, MAR. 2024.

\bibitem{wang2024robust}
Y.~Wang, X.~Yi, H.~Hou, W.~Wang, and S.~Jin, ``Robust symbol-level precoding
  for massive {MIMO} communication under channel aging,'' \emph{IEEE Trans.
  Wireless Commun.}, vol.~23, no.~9, pp. 10\,864--10\,878, MAY 2024.

\bibitem{zhu2024joint}
Y.~Zhu, J.~Zhuang, G.~Sun, H.~Hou, L.~You, and W.~Wang, ``Joint channel
  estimation and prediction for massive {MIMO} with frequency hopping
  sounding,'' \emph{IEEE Trans. Wireless Commun.}, 2024.

\bibitem{luo2024user}
X.~Luo, X.~Lu, B.~Jin, B.~Yin, and K.~Yang, ``User grouping and resource
  allocation for joint communication and positioning in {mmWave} multi-cell
  networks,'' \emph{IEEE Trans. Wireless Commun.}, vol.~23, no.~10, pp.
  14\,096--14\,108, FEB. 2024.

\bibitem{price2007communication}
R.~Price and P.~E. Green, ``A communication technique for multipath channels,''
  \emph{Proc. IRE}, vol.~46, no.~3, pp. 555--570, JAN. 2007.

\bibitem{Zeng2024}
Y.~Zeng, J.~Chen, J.~Xu, D.~Wu, X.~Xu, S.~Jin, X.~Gao, D.~Gesbert, S.~Cui, and
  R.~Zhang, ``A tutorial on environment-aware communications via channel
  knowledge map for {6G},'' \emph{IEEE Commun. Surveys Tuts.}, vol.~26, no.~3,
  pp. 1478--1519, 2024.

\bibitem{jafri2022robust}
M.~Jafri, A.~Anand, S.~Srivastava, A.~K. Jagannatham, and L.~Hanzo, ``Robust
  distributed hybrid beamforming in coordinated multi-user multi-cell {mmWave}
  {MIMO} systems relying on imperfect {CSI},'' \emph{IEEE Trans. Commun.},
  vol.~70, no.~12, pp. 8123--8137, 2022.

\bibitem{GiordanoWCNC2018SRScoord}
L.~Galati~Giordano, L.~Campanalonga, D.~Lopez-Perez, A.~Garcia-Rodriguez,
  G.~Geraci, P.~Baracca, and M.~Magarini, ``Uplink sounding reference signal
  coordination to combat pilot contamination in 5g massive {MIMO},'' in
  \emph{Proc. IEEE Wireless Commun. Netw. Conf. (WCNC)}, APR. 2018, pp. 1--6.

\bibitem{zhuang2025dmrs}
J.~Zhuang, H.~Hou, M.~Tang, W.~Wang, S.~Jin, and V.~K. Lau, ``{DMRS}-based
  uplink channel estimation for {MU-MIMO} systems with location-specific {SCSI}
  acquisition,'' \emph{arXiv preprint arXiv:2506.11899}, 2025.

\bibitem{jaeckel2018efficient}
S.~Jaeckel, L.~Raschkowski, F.~Burkhardt, and L.~Thiele, ``Efficient
  sum-of-sinusoids-based spatial consistency for the {3GPP} new-radio channel
  model,'' in \emph{Proc. IEEE Global Commun. Conf. Workshops (GC
  Wkshps)}.\hskip 1em plus 0.5em minus 0.4em\relax IEEE, DEC. 2018.

\bibitem{zhu20213gpp}
Q.~Zhu, C.-X. Wang, B.~Hua, K.~Mao, S.~Jiang, and M.~Yao, ``{3GPP TR 38.901
  Channel Model},'' in \emph{{Wiley} {5G} {Ref}: {Essential} {5G} {Reference}
  Online}.\hskip 1em plus 0.5em minus 0.4em\relax Hoboken, NJ, USA: Wiley, JAN.
  2021, pp. 1--35.

\bibitem{jafri2024asynchronous}
M.~Jafri, S.~Srivastava, S.~Kumar, A.~K. Jagannatham, and L.~Hanzo,
  ``Asynchronous distributed coordinated hybrid precoding in multi-cell
  {mmWave} wireless networks,'' \emph{IEEE Open J. Veh. Technol.}, vol.~5, pp.
  200--218, 2024.

\bibitem{wang2022channel}
L.~Wang, G.~Liu, J.~Xue, and K.-K. Wong, ``Channel prediction using ordinary
  differential equations for {MIMO} systems,'' \emph{IEEE Trans. Veh.
  Technol.}, vol.~72, no.~2, pp. 2111--2119, 2022.

\bibitem{bjornson2019making}
E.~Bj{\"o}rnson and L.~Sanguinetti, ``Making cell-free massive {MIMO}
  competitive with {MMSE} processing and centralized implementation,''
  \emph{IEEE Trans. Wireless Commun.}, vol.~19, no.~1, pp. 77--90, 2019.

\bibitem{jafri2024distributed}
M.~Jafri, S.~Kumar, S.~Srivastava, and A.~K. Jagannatham, ``Distributed hybrid
  beamforming in {mmWave} multi-cell systems in the presence of cell-edge users
  relying on stochastic channel uncertainty,'' \emph{IEEE Trans. Commun.},
  2024.

\bibitem{jaeckel2014quadriga}
S.~Jaeckel, L.~Raschkowski, K.~B{\"o}rner, and L.~Thiele, ``{QuaDRiGa}: A {3-D}
  multi-cell channel model with time evolution for enabling virtual field
  trials,'' \emph{IEEE Trans. Antennas Propag.}, vol.~62, no.~6, pp.
  3242--3256, NOV. 2014.

\bibitem{3gpp2018study}
{3GPP}, ``Study on channel model for frequencies from 0.5 to 100 {GHz},''
  \emph{3GPP Tech. Spec. 38.901}, JAN. 2018.

\end{thebibliography}

\end{document}